\documentclass[letterpaper,twocolumn,10pt]{article}

\usepackage[hyphens]{url}
\usepackage{usenix}
\usepackage{tikz}
\usepackage{amsmath}
\usepackage{enumitem}
\usepackage{graphicx}

\begin{document}

\date{}

\title{\Large \bf Deep Motion Masking for Secure, Usable, and Scalable\\Real-Time Anonymization of Virtual Reality Motion Data}

\author{
{\rm ~ ~ ~ ~ ~ ~ ~ ~ ~ ~ ~}\\
~ ~ ~ ~ ~ ~ ~ ~ ~ ~ ~ ~
\and
{\rm Vivek Nair}\\
UC Berkeley
\and
{\rm Wenbo Guo}\\
Purdue University
\and
{\rm James F. O’Brien}\\
UC Berkeley
\and
{\rm ~ ~ ~ ~ ~ ~ ~ ~ ~ ~ ~}\\
~ ~ ~ ~ ~ ~ ~ ~ ~ ~ ~ ~
\and
{\rm Louis Rosenberg}\\
Unanimous AI
\and
{\rm Dawn Song}\\
UC Berkeley
} 

\maketitle

\begin{abstract}
Virtual reality (VR) and ``metaverse'' systems have recently seen a resurgence in interest and investment as major technology companies continue to enter the space.
However, recent studies have demonstrated that the motion tracking ``telemetry'' data used by nearly all VR applications is as uniquely identifiable as a fingerprint scan, raising significant privacy concerns surrounding metaverse technologies.
Although previous attempts have been made to anonymize VR motion data, we present in this paper a state-of-the-art VR identification model that can convincingly bypass known defensive countermeasures.
We then propose a new ``deep motion masking'' approach that scalably facilitates the real-time anonymization of VR telemetry data.
Through a large-scale user study ($N=182$), we demonstrate that our method is significantly more usable and private than existing VR anonymity systems.
\end{abstract}

\section{Introduction}
\label{sec:introduction}

The recent resurgence of research and development investiture into virtual reality (VR) and ``metaverse'' technologies has created an accelerated pace of technological improvements that are steadily making their way to consumer-facing VR devices.
Newly announced products like the Apple Vision Pro \cite{visionpro} and Meta Quest 3 \cite{quest3} blur the lines between virtual and augmented reality, resulting in extended reality (XR) systems that are expected to be more deeply and seamlessly integrated with our daily lives than ever before.
Despite these changes, motion capture ``telemetry'' data remains fundamental to the operation of nearly all XR devices and applications.

While human motion patterns have been recognized as a uniquely identifiable and revealing biometric since at least the 1970s \cite{cutting_recognizing_1977, kozlowski_recognizing_1977}, researchers are only beginning to understand the implications of this for motion data captured by XR devices. Recent studies have demonstrated that head and hand motion data captured by a VR device can be used to uniquely identify its user across a variety of applications \cite{miller_personal_2020, schell_comparison_2022, 10027854}, over long periods of time \cite{miller2023largescale, rack2023extensible}, and at a rate of over 1 in 50,000 \cite{291259}, comparable to that of a fingerprint scan \cite{nist}.
Moreover, they show that a variety of potentially sensitive user data attributes can be inferred directly from VR telemetry streams \cite{nair2023inferring}.
Such results raise serious questions about whether XR devices can be used without involuntarily revealing a plethora of personal information to the device, application, and other XR users.

Researchers have proposed a number of methods for anonymizing VR motion data without unduly degrading the user experience  \cite{miller_personal_2020, 9583839, nair2023going}.
However, current anonymization methods underestimate the identifiability of motion data when using sophisticated models trained on large datasets. In this paper, we present a best-in-class VR identification model that achieves over $90\%$ cross-session identification accuracy with 500 users, even when using existing countermeasures.
We then propose ``deep motion masking,'' a technique that uses deep learning to effectively anonymize VR motion data.

Deep motion masking represents a multi-axis improvement over prior VR anonymization methods. Through a comprehensive evaluation, we demonstrate a $2.7 \times$ improvement in the indistinguishability of anonymized motion data, and an over $20 \times$ improvement in cross-session unlinkability. Our proposed system is capable of low-latency real-time anonymization of VR telemetry streams, making it concretely practical for deployment in new and existing VR systems.\\

\vspace{-0.2em}
\noindent \textbf{Contributions:}
\begin{itemize}[leftmargin=*]
    \itemsep 0em
    \item We present a new, state-of-the-art VR identification model that can bypass existing VR anonymity systems (\S\ref{sec:motivation}).
    \item We propose a ``deep motion masking'' technique for scalable, real-time anonymization of VR telemetry data (\S\ref{sec:method}).
    \item Using new and existing VR identification models, our evaluation ($N=1,000$ users) shows at least a $20\times$ improvement in anonymity over prior VR privacy approaches (\S\ref{sec:anonymity}).
    \item Our large-scale usability study ($N=182$ participants) demonstrates a nearly $3\times$ improvement in the indistinguishability of resulting anonymized motion data (\S\ref{sec:usability}).
    \item In simulations, we show that our anonymizer has minimal impact on perceived interactions with virtual objects (\S\ref{sec:interactivity}).
\end{itemize}

\eject
\section{Background}
\label{sec:background}

Virtual reality systems use a variety of input and output devices to create an immersive visual, auditory, and haptic experience for users.
However, in addition to being used for its intended purposes, the data generated by VR device sensors can be used adversarially to infer private user information.

The SoK of Garrido et al. \cite{garrido2024sok} provides a standard information flow and threat model for VR privacy research. In this section, we briefly describe the typical information flow of motion data generated by a VR device. We then recount the threat model of Garrido et al. so as to position our work within the broader landscape of VR privacy research.

\subsection{Information Flow}
\label{sec:informationflow}

A typical VR system sold today includes one head-mounted display (HMD) and two hand-held controllers. At a rate of between 60 and 144 times per second, the VR device measures the position and orientation of each of these three devices in 3D space (with six degrees of freedom), creating a ``telemetry stream.'' 
These measurements are typically generated using a combination of inertial measurement units (IMUs) and either onboard cameras (also known as ``inside-out'' tracking) or external tracking stations (known as ``outside-in'' tracking).

In addition to motion tracking, many modern VR devices contain a number of additional sensors, such as LIDAR arrays, microphones, cameras, eye tracking, and body tracking systems. However, the focus of this paper is on the basic head and hand motion telemetry data that remains universal and fundamental to nearly all VR devices and applications.

In a typical VR system, telemetry data is generated by the VR device hardware and is then consumed by a VR application via an API provided by the device's firmware. The VR application uses this data to render frames to be displayed on the VR device, as well as to generate auditory and haptic stimuli for the user. In the case of a multi-user or ``metaverse'' application, the telemetry data is also forwarded to a server, which in turn forwards the data to other users in order to render a virtual representation (or ``avatar'') of the user on the devices of other users in the same virtual environment.

\subsection{Threat Model}
\label{sec:threatmodel}

Because each entity in the above information flow (namely, the VR hardware, the application, the server, and another user) has access to the motion data stream of a target user, they could all potentially misuse such data in order to infer private user information. As such, they are all considered potential adversaries in the Garrido et al. threat model.

\begin{figure}[h]
    \centering
    \includegraphics[width=0.75 \linewidth]{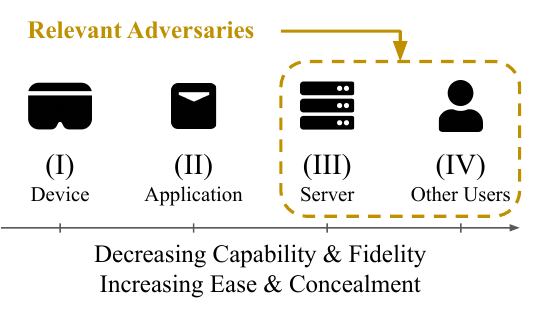}
    \caption{VR privacy threat model and relevant adversaries.}
    \label{fig:threatmodel}
\end{figure}

Figure \ref{fig:threatmodel} illustrates the threat model of Garrido et al. and indicates the adversaries relevant to this paper. As in prior work, our emphasis in this paper is on protecting the motion data visible to external adversaries, namely VR game servers and other VR users. These adversaries are considered ``weaker'' in the Garrido et al. threat model, meaning that attacks available to them are typically available to all other adversaries. Moreover, attacks performed by these adversaries are generally the hardest to detect due to their remote and decentralized nature.

In summary, the focus of this paper is on the threat posed by broadcasting head and hand motion data to servers and external users in multi-user VR applications. These threats are amongst the most realistic, universal, and pernicious security and privacy challenges present in VR devices today.
\section{Related Work}
\label{sec:relatedwork}

Security and privacy in XR is a rapidly growing area of research that is summarized well by a number of existing survey and position papers \cite{garrido2024sok, priv_implications_VR, 10.1145/2580723.2580730, nair2023truth, giaretta2022security, priv_metaverse, metaverse_privacy_1, de_guzman_security_2020}. In this section, we summarize the body of work most directly relevant to this paper. We begin by detailing the history of motion-based biometrics and describe a number of studies illustrating relevant attacks on VR motion data. We then outline the relatively small number of proposed countermeasures to said attacks in comparison with the defensive system proposed herein.

\subsection{Motion Biometrics}
\label{sec:motionbiometrics}

Since at least the 1970s, researchers have shown that individuals reveal a plethora of information about themselves via their motion. In 1977, Cutting and Kozlowski demonstrated that participants in a laboratory study could identify their friends just by viewing the motion of eight tracked points affixed to the body \cite{cutting_recognizing_1977}. Later that year, they showed that the gender of the participants could be identified by a stranger with statistically significant accuracy \cite{kozlowski_recognizing_1977}. More recently, Jain et al. \cite{jain_is_2016} found that the age of an individual can also be accurately determined from their motion patterns alone.

The motion capture data produced by VR devices is remarkably similar to the motion data used in the above studies, although it is somewhat distinct in that only three tracked locations are typically observed in VR rather than eight or more. For this reason, a new line of research has emerged to examine whether VR motion data reveals private user information.

\subsection{VR Attacks}
\label{sec:vrattacks}

Prior work researching the privacy consequences of VR motion data specifically may be broadly categorized into identification studies, which use VR motion data to uniquely identify VR users, and profiling studies, which instead attempt to infer specific attributes such as age and gender.

Many papers have analyzed the possibility of motion-based identification in VR, which are summarized well by the SoK papers of Stephenson et al. \cite{9833742} and Garrido et al. \cite{garrido2024sok}. Some key results in this field are summarized in Table \ref{tab:ident}.

\begin{table}[h]
\centering
\resizebox{\columnwidth}{!}{%
\begin{tabular}{|l|l|l|l|l|}
\hline
\textbf{Study} & \textbf{Algorithm} & \textbf{Activity} & \textbf{Users} & \textbf{Accuracy} \\ \hline
Kupin et al. (2019) \cite{kupin} & KNN & Ball Throwing & 14 & 93\% \\ \hline
Pfeuffer et al. (2019) \cite{pfeuffer} & Random Forest & Interactive Tasks & 22 & 40\% \\ \hline
Miller et al. (2020) \cite{miller_personal_2020} & Random Forest & 360$^{\circ}$ Videos & 511 & 95\% \\ \hline
Miller et al. (2020) \cite{miller_2020} & KNN & Ball Throwing & 46 & 97\% \\ \hline
Liebers et al. (2021) \cite{10.1145/3411764.3445528} & LSTM & Archery Game & 16 & 90\% \\ \hline
Miller et al. (2021) \cite{miller_2021} & MLP & Ball Throwing & 46 & 99\% \\ \hline
Moore et al. (2021) \cite{9583839} & Random Forest & Training Application & 60 & 91\% \\ \hline
Tricomi et al. (2022) \cite{10027854} & Logistic Regression & Robot Teleoperation & 30 & 95\% \\ \hline
Liebers et al. (2022) \cite{liebers2022} & Random Forest & Bowling Game & 16 & 95\% \\ \hline
Nair et al. (2023) \cite{291259} & LightGBM & Rhythm Game & 55,541 & 94\% \\ \hline
Rack et al. (2023) \cite{rack2023extensible} & CNN & FPS Game & 63 & 98\% \\ \hline
Miller et al. (2023) \cite{miller2023largescale} & Random Forest & Virtual Classroom & 232 & 70\% \\ \hline
Liebers et al. (2023) \cite{liebers23} & Random Forest & Rhythm Game & 15 & 71\% \\ \hline
\end{tabular}%
}
\caption{Notable prior VR motion identification studies.}
\label{tab:ident}
\end{table}

Overall, a large number of studies have concluded that the head and hand motion data captured by VR devices is capable of accurately identifying users in a variety of applications.

A second major class of VR motion privacy research investigates profiling specific user attributes from head and hand movement patterns. For example, Tricomi et al. \cite{10027854} use eye tracking data in addition to head and hand motion to accurately infer the gender and age of about 35 VR users.
More recently, in a study of 1,006 VR users, Nair et al. \cite{nair2023inferring} demonstrated that over 40 personal attributes, ranging from background and
demographics to behavioral patterns and health information, can be accurately and consistently inferred from VR motion data alone. Additionally, multiple studies have demonstrated that adversarially designed VR applications can harvest further user data than passive observation alone \cite{9319051, nair_exploring_2023}.

In summary, while motion data is an essential part of most VR experiences, the bulk of prevailing research indicates that sharing this data with third parties carries significant security and privacy consequences. As such, it is imperative to develop systems that enhance the privacy of VR motion data without impeding essential VR application functionality.

\subsection{VR Defenses}
\label{sec:vrdefenses}

There are a number of fundamental challenges that complicate the development of privacy-preserving mechanisms for VR motion data. First, there is a lack of fine-grained access control, as the exact same telemetry data that is necessary to provide legitimate multi-user functionality can also be used for adversarial purposes.
While related work proposes access control for environmental data in XR systems \cite{291287, 180378}, the equivalent does not yet exist for motion data.
Thus, instead of eliminating access to the VR motion stream, the data must somehow be transformed such that potential adversarial uses are thwarted while legitimate functionality remains intact. 
We compare this objective to a real-time voice changer that makes a speaker's voice unrecognizable while preserving spoken content and producing natural-sounding speech \cite{285349}.

Further complicating attempts to protect the privacy of VR motion data is the need for any resulting defensive system to be real-time and low-latency. In many cases, even slight delays in a VR rendering pipeline can result in a phenomenon known as ``VR sickness'' \cite{10.1145/333329.333344}. This means that any realistic countermeasure must be fast and respect causality (i.e., cannot use future data to process past data). By contrast, VR attackers can be slow and non-causal, using an entire session of motion data at once to conduct their attack, creating a fundamental imbalance between attacker and defender capabilities in VR. Many adjacent research areas, such as gait recognition, lack these constraints. As such, proposed defenses in related domains are not necessarily directly applicable to VR.

With respect to VR motion data specifically, Miller et al. \cite{miller_personal_2020} have suggested a motion transmission method that only communicates joint rotation data, resulting in a 75\% reduction in identifiability. Similarly, Moore et al. \cite{9583839} suggest a method that transmits velocity data rather than positions, observing a 57\% reduction in identification accuracy. On the contrary, Rack et al. \cite{rack2023extensible} actually recommend the use of body-relative velocity and acceleration for identification purposes, citing an increase in identification accuracy rather than a reduction.

Most notably, MetaGuard \cite{nair2023going} anonymizes \footnote{We use ``anonymize'' and ``deidentify'' synonymously in this paper.} VR motion data by applying bounded Laplacian noise \cite{boundedtruncatedIBM} to specific dimensions of the telemetry stream that correspond to identifiable anthropometrics like height and wingspan.
As a result, the system satisfies $\varepsilon$-differential privacy \cite{dwork_algorithmic_2013} and theoretically achieves an optimal noise versus privacy trade-off.

We seek to improve upon the existing countermeasures for two major reasons. First, the ad-hoc nature of the dimensions selected for anonymization is unlikely to be scalable when additional tracked locations are introduced to the system.
As full-body tracking systems are increasingly becoming the norm for new VR devices \cite{quest3, visionpro}, proposed defensive mechanisms should at least plausibly demonstrate the potential to scale to more than three tracked locations in the future.

Furthermore, as demonstrated in \S\ref{sec:motivation}, VR identification models can be far more powerful than previously imagined if enough training data is available. Existing countermeasures did not anticipate the extent to which users may be identified from a reduced set of features given a sufficiently powerful model.
In the following section, we describe the substantial dataset utilized in this study, which enabled significant improvements in both offensive and defensive VR technologies.
\eject

\section{Dataset}
\label{sec:dataset}

The primary source of motion capture data used in this paper is ``Beat Saber,'' a VR rhythm game. This section briefly describes and motivates our choice of application and dataset.

\subsection{Beat Saber}
\label{sec:beatsaber}

\begin{figure}[h]
    \centering
    \includegraphics[width=\linewidth]{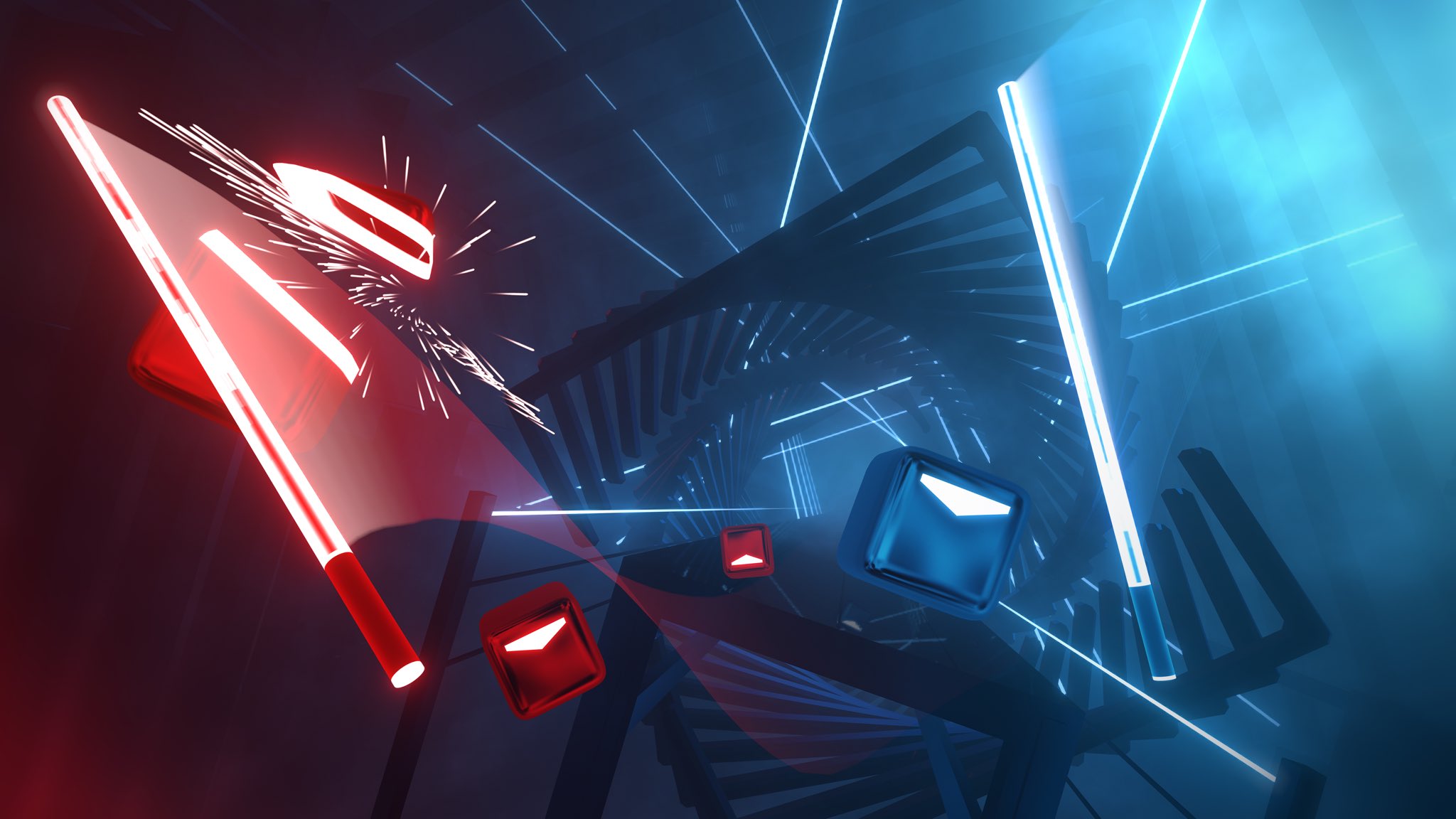}
    \caption{``Beat Saber,'' a VR rhythm game.}
    \label{fig:beatsaber}
\end{figure}

Figure \ref{fig:beatsaber} shows a first-person view of ``Beat Saber,'' \cite{beat_saber} a popular VR rhythm game in which players use a pair of sabers held in each hand to slice flying blocks that represent musical beats. Beat Saber is the most popular and highest-grossing VR game of all time \cite{wobbeking_beat_2022}, making it a representative example of a non-adversarial VR game with multi-player functionality.

Beat Saber is split into a number of levels or ``maps,'' which consist of an audio track (typically a song) and a series of in-game obstacles that players must accurately interact with to achieve a high score. 
Users around the world can play and compete to achieve the highest possible score on hundreds of official maps and thousands of unofficial user-created maps.

\subsection{BeatLeader}
\label{sec:beatleader}

``BeatLeader'' \cite{beatleader} is an open-source third-party leaderboard website for Beat Saber. Beat Saber players may opt-in to use BeatLeader by installing a custom extension to the base game.
After playing a Beat Saber level with the extension installed, scores are automatically uploaded to an online leaderboard.

When uploading a score to BeatLeader, a recording of the user’s
motion telemetry during play is automatically captured and attached
to their submission. The score and corresponding recording are then made publicly available on the BeatLeader website to demonstrate the authenticity of the submission.

\subsection{BOXRR-23}
\label{sec:boxrr23}

BOXRR-23 \cite{nair2023berkeley} is a publicly available VR motion dataset 
that contains aggregated anonymized data from upstream sources such as BeatLeader. It contains over 3.5 million VR motion capture recordings, submitted by nearly 100,000 users between February 2022 and April 2023.
BOXRR-23 also contains data from ScoreSaber \cite{scoresaber} and PolyGone \cite{polygone}; however, only the BeatLeader portion of the data is used in this paper.

Our motivation for selecting this dataset is threefold. First, BOXRR-23 is multiple orders of magnitude larger than the next largest VR motion dataset, making it an obvious choice for training deep learning models.
Additionally, the authors explicitly endorse using the dataset for security and privacy research, and state that the dataset underwent stringent ethical and legal review for those purposes prior to its release.
Finally, using an already-public dataset will improve the transparency, reproducibility, and extensibility of this work.
\section{Motivation}
\label{sec:motivation}

We now present a series of introductory experiments on motion-based identification in VR using the dataset of \S\ref{sec:dataset}.
We describe the basic principles behind existing VR identification models and then show that with a sufficiently large volume of data, models can be trained that are far more robust and capable than those discussed in prior work. The aim of this section is not to serve as the main contribution of this paper but rather to motivate our new defensive approach by demonstrating the insufficiency of existing countermeasures.

\subsection{Prevailing Architectures}
\label{sec:prevailingarchitectures}

At present, most existing papers on VR user identification utilize classical machine learning models, such as those based on the Random Forest \cite{breiman2001random} and LightGBM \cite{ke_lightgbm_2017} architectures.
The motivation for using these models over theoretically more powerful deep learning approaches is that deep learning typically requires a significantly larger volume of data to successfully train and converge, whereas tree-based architectures can produce generalizable classifiers with fewer samples per user.

On the other hand, the sequential time-series format of VR motion data streams is not a natural fit for tree-based models, which usually require a one-dimensional tabular data format.
As such, prior works suggest deliberate feature engineering to convert motion data streams into tabular samples by using summary statistics to eliminate the time dimension.

Specifically, Pfeuffer et al. \cite{pfeuffer} suggest dividing motion data into one-second chunks, and then converting each chunk into a flat feature vector by taking four statistics (min, max, mean, and standard deviation) across each tracked dimension. Miller et al. \cite{miller_2020} use a very similar approach, but also include the median of each axis. Moore et al. \cite{9583839} use identical features to Miller, while Nair et al. \cite{291259} use similar features but add contextual data specific to the VR application. At a high level, many prior works have found the basic idea of summarizing one-second chunks of motion to be highly effective.

\eject

Surprisingly, the method of using one-second summary statistics has in some instances outperformed sequential deep learning models even when sufficiently large datasets are present.
For example, Nair et al. \cite{291259} report that LightGBM with tabular summary statistics outperformed MLP, GRU, and LSTM models despite using a fairly large amount of data.

For reasons yet unknown, the basic notion of summarizing one-second subsequences of larger motion recordings seems uniquely well-suited for identifying VR users. Thus, we are motivated to replicate this approach using deep learning architectures in order to achieve better identification performance.

\subsection{LSTM Funnel Architecture}
\label{sec:lstmfunnelarchitecture}

\begin{figure}[h]
    \centering
    \includegraphics[width=\linewidth]{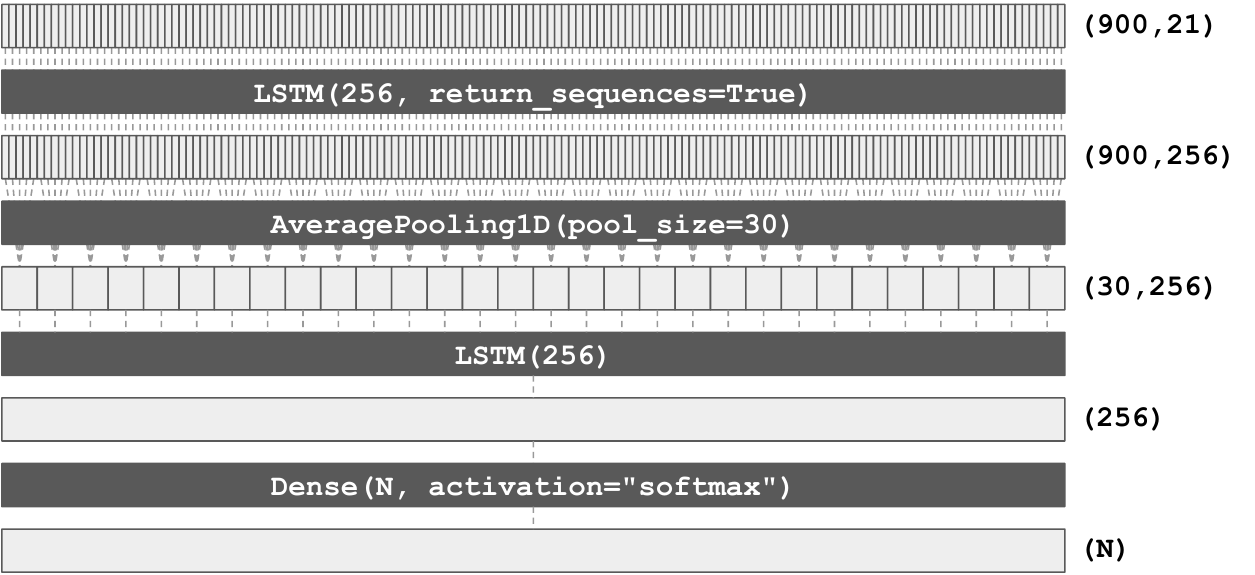}
    \caption{``LSTM funnel'' identification architecture.}
    \label{fig:funnel}
\end{figure}

In this section, we propose a new deep learning architecture that aims to internally replicate the idea of summarizing one-second motion subsequences by using a combination of Long Short-Term Memory (LSTM) \cite{hochreiter1997long} and Multi-Layer Perceptron (MLP) \cite{goodfellow2016deep} layers.
Figure \ref{fig:funnel} illustrates how the proposed architecture may be used to identify VR motion sequences. The model receives as input a 30-second motion sequence normalized to 30 frames per second, thus containing 900 frames in total. Using an LSTM layer, each frame is converted into a 256-dimensional feature vector. Then, an average pooling layer combines each one-second (30-frame) subsequence into a 256-dimensional summary. Next, another LSTM layer combines the sequence of 30 256-dimensional summaries into a flat 256-dimensional embedding. Finally, a fully connected MLP layer with softmax activation produces a classification output, with optional additional dense layers in between.

In essence, the architecture described above continues to represent VR motion sequences using summary statistics taken across one-second chunks, yet is able to outperform prior approaches for a few major reasons. 
First, instead of manually specifying summary statistics to be taken, such as mean, standard deviation, etc., the model is allowed to learn its own relevant statistics via the first LSTM layer.
Second, instead of manually specifying how to summarize the classification of each subsequence, such as via a logarithmic sum of probabilities, the model is allowed to learn its own meta-classification method via the second LSTM and subsequent MLP layers. Moreover, the ``featurization'' and ``classification'' parts of the model are trained together in an end-to-end fashion, allowing the model to learn how to create complex statistics that result in optimal classification results.

We term this approach the ``LSTM funnel'' architecture due to the dimensionality reduction performed by the average pooling layer.
While fairly simple overall, to the best of our knowledge, this architecture has not yet been disclosed in general or has not been used for similar purposes.

\subsection{Worst-Case Identifiability}
\label{sec:worstcaseidentifiability}

We now demonstrate how the LSTM funnel architecture can be used to drive significant improvements in motion-based identification accuracy, provided a large amount of training data per user is available.
Using the dataset of \S\ref{sec:dataset}, we first found the 500 users for which the greatest number of individual recordings were available. For these top 500 users, an average of 821 recordings were available per user, with each recording averaging about three minutes in length.
We used the 500 most recent recordings of each user for our evaluation, with 400 of these recordings being used for training, 50 for validation, and the remaining 50 being used for testing.
To conform to the architecture of \S\ref{sec:lstmfunnelarchitecture}, only the first 30 seconds of each recording were utilized, and recordings were normalized to a constant 30 frames per second by using a numerical linear interpolation for positional coordinates and a spherical linear interpolation for orientation quaternions.

To evaluate the performance of the LSTM funnel architecture on this particular dataset,  we implemented the architecture of Figure \ref{fig:funnel} in Keras v2.10.1 \cite{keras} and trained it for 500 epochs on the described dataset using the Adam optimizer \cite{kingma2014adam} with a learning rate of 0.001.
The validation dataset was used for early stopping after 25 epochs of no improvement. For the sake of comparison, we also trained and tested several previously proposed identification model architectures using the same dataset, the results of which were as follows:

\begin{itemize}[leftmargin=*]
    \item Our new LSTM funnel architecture achieves a per-sample accuracy of 98.12\% and a per-user accuracy of 100.00\%.
    \item The Nair et al. \cite{291259} architecture achieves a per-sample accuracy of 71.66\% and a per-user accuracy of 100.00\%.
    \item The Miller et al. \cite{miller2023largescale} architecture achieves a per-sample accuracy of 56.59\% and a per-user accuracy of 97.60\%.
\end{itemize}

As evidenced by the above results, our architecture substantially exceeds the identification performance of the most notable prior models when using identical datasets.
This, on its own, is not entirely surprising, given that we used over three hours of training data per user to perform this demonstration, which also exceeds all prior works; the previously proposed models and featurization approaches were not designed to take full advantage of this volume of data. 
However, the robustness of our new architecture to reductions in input dimensionality is, to our knowledge, unprecedented:

\begin{itemize}[leftmargin=*]
    \itemsep 0em
    \item The original representation with the full 21 features ($\{\mathit{head},\mathit{left\_hand},\mathit{right\_hand}\}\times\{x,y,z,i,j,k,w\}$) gives a sample accuracy of 98.12\% and a user accuracy of 100.00\%.
    \item Removing the head, the remaining 14 features ($\{\mathit{left\_hand},\mathit{right\_hand}\}\times\{x,y,z,i,j,k,w\}$) reduce sample accuracy to 94.76\% (and still 100\% user accuracy).
    \item Using only hand rotations, the remaining 8 features ($\{\mathit{left\_hand},\mathit{right\_hand}\}\times\{i,j,k,w\}$) give a sample accuracy of 93.42\% and a user accuracy of 100.00\%.
    \item Using only left hand rotations, the remaining 4 features ($\{\mathit{left\_hand}\}\times\{i,j,k,w\}$) still result in a sample accuracy of 92.77\% and a user accuracy of 100.00\%.
    \item Using only left hand rotational magnitude, the single feature ($\{\mathit{left\_hand}\}\times\{w\}$) still results in a sample accuracy of 84.23\% and a user accuracy of 100.00\%.
\end{itemize}

In other words, by observing just the absolute magnitude of the rotation of one hand of a user for a period of just 30 seconds, the model can still correctly identify the user out of 500 options with nearly 85\% accuracy, provided it was first trained on over 3 hours of data for each user.

Today, obtaining 200 minutes of motion capture data for a user may seem like an absolute worst-case scenario from a privacy perspective, with the 500 individuals used in our demonstration perhaps being amongst the only individuals in the world for which this amount of data is readily accessible.
However, if extended reality truly replaces existing mobile devices as a default method of human-computer interaction for millions of users in the near future, having multiple hours of cumulative time spent using XR devices may soon come to represent an average or even below-average usage pattern.

\subsection{Prevailing Defenses}
\label{sec:prevailingdefenses}

In light of the new findings discussed above, we now briefly revisit and reevaluate the existing proposals for countermeasures against motion-based identification in VR:

\begin{itemize}[leftmargin=*]
    \itemsep 0em
    \item Miller et al. \cite{miller_personal_2020} have suggested transmitting only certain rotational dimensions rather than positional data. However, as demonstrated by the results of \S\ref{sec:worstcaseidentifiability}, hand rotation values alone are now sufficient to accurately deanonymize users.
    \item Moore et al. \cite{9583839} suggest transmitting velocity data rather than positions. However, one can recover rotational magnitude by integrating angular velocities, which we have shown is sufficient for identification. Others have found that joint velocities are actually more identifiable than positions \cite{schell_comparison_2022}.
    \item Nair et al. \cite{nair2023going} suggest using differential privacy to randomize particular anthropometric measurements like height and wingspan. This method has no impact on rotation values, which we have shown are sufficient to deanonymize users.
\end{itemize}

Each of the existing countermeasures was not designed with the understanding that any individual axis of motion data could be sufficient on its own to deanonymize users if a large enough amount of training data is utilized. 
With this in mind, a truly effective solution must comprehensively anonymize every individual axis present in the motion telemetry stream, as well as all of the identifiable relationships between those dimensions.
Manually engineering an adequate solution for each dimension is already on the edge of feasibility with the 21 dimensions tracked by current systems, and becomes completely impractical when given the hundreds of dimensions measured by next-generation full-body tracking systems. Therefore, we are motivated to investigate the use of deep learning to comprehensively anonymize VR telemetry data and construct a more scalable motion anonymization system.

\subsection{Problem Statement}
\label{sec:problemstatement}

Having motivated our reasons for wanting to improve VR anonymization techniques beyond the current state of the art, we present in this paper a new ``deep motion masking'' approach to VR motion anonymization, which we use to create an improved motion anonymization system.
The goals of our new system and approach are as follows:

\begin{itemize}[leftmargin=*]
    \itemsep 0em
    \item \textbf{Anonymity}: The primary goal of the system is to prevent users from being identified based on their motion data. Specifically, we invoke the same notion of anonymity as used in MetaGuard \cite{nair2023going}, \textit{cross-session unlinkability}; given motion data with known user identities in a first session, the adversaries relevant to this paper (see \S\ref{sec:threatmodel}) should not be able to identify the same set of users using their anonymized motion data from a second session.
    As in MetaGuard, we assume that adversaries have no other means of linking participant identities across sessions, such as IP addresses.
    \item \textbf{Usability}: The system must not significantly degrade the user experience by anonymizing user motion data.
    Specifically, we target the strong notion of \textit{indistinguishability} of anonymized motion data from unmodified VR motion data.
\end{itemize}

We contend that these properties are both necessary and sufficient for a practical VR motion privacy system. 
Clearly, anonymity is a necessary property of a motion privacy system in order to protect the identity of VR users. In particular, the cross-session unlinkability definition we use prevents adversaries from tracking users from one usage context to another and aggregating an increasingly detailed profile of the user over time.
Of course, as discussed in \S\ref{sec:vrattacks}, known VR attacks go beyond the identification of users, and include the ability to profile various personal attributes. However, if anonymized, such attributes will no longer be linkable to the identity of a particular user. Further, a system that is effective at anonymizing users must, in practice, also effectively obscure any set of personal attributes that can be correlated to their identity.

\eject

Similarly, the usability of the resulting system is sufficiently ensured by the indistinguishability of anonymized motion data, as anonymized motion data that is indistinguishable from unmodified natural human motion data cannot negatively impact the user experience.
If the anonymized motion data diminished the usability of the VR system in any way, it would, in fact, be distinguishable from unmodified human motion data by virtue of causing said diminution.

In addition to the main properties described above, we note two further ``soft'' requirements that influenced our design choices. While these properties are technically already encapsulated in the above goals, they serve to further constrain the design of our system and to distinguish its capabilities from those of previous defensive systems like MetaGuard \cite{nair2023going}:

\begin{itemize}[leftmargin=*]
    \itemsep 0em
    \item \textbf{Scalability}: The anonymization system should comprehensively anonymize every axis of motion data without manually engineering a solution for each feature.
    \item \textbf{Interactivity}: The system should minimize the perceived impact of the anonymization process on the interaction of the user with objects in the virtual world.
\end{itemize}

With these properties in mind, we now describe our new proposal for a ``deep motion masking'' system.
\section{Method}
\label{sec:method}

At a high level, our method involves decomposing the plausible variance of human motion sequences into action-related variance and user-related variance. For this purpose, we train an ``action encoder'' model, which learns an embedding for the action a user is taking while ignoring the user's identity, and a ``user encoder'' model, which learns an embedding for the user's identity while ignoring the action they are taking.
We then train an ``anonymizer'' model that anonymizes motion sequences by changing their user embedding without changing their action embedding.
Finally, we train a ``normalizer'' model to remove unwanted noise added by the anonymizer.

Each of the models we describe was implemented in Keras \cite{keras} and trained using the Adam optimizer \cite{kingma2014adam} with a diminishing learning rate scheduler and early stopping based on a validation set.
For each training step, and throughout this paper, we provide benchmarking results in \S\ref{app:benchmarking}.

\subsection{Action Similarity}
\label{sec:actionsimilarity}

\begin{figure}[h]
    \centering
    \includegraphics[width=\linewidth]{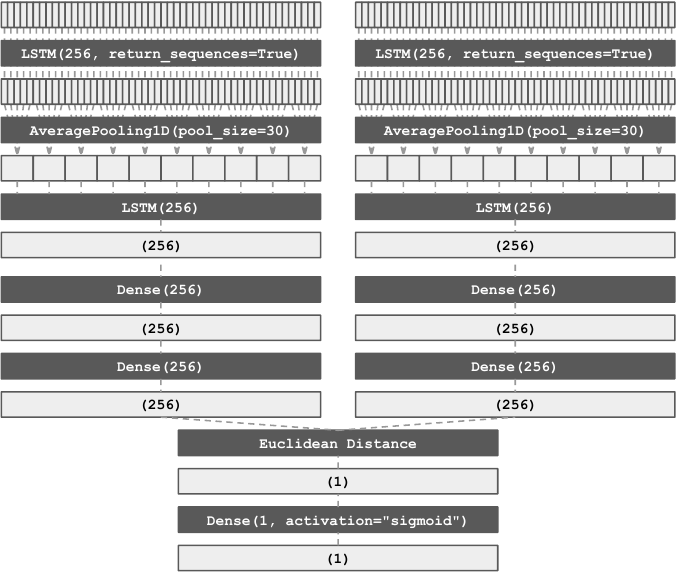}
    \caption{Siamese architecture for similarity models.}
    \label{fig:similarity}
\end{figure}

First, we describe our method for measuring the similarity of the ``action'' performed in two separate VR motion sequences. To achieve this, we train an ``action similarity'' model using the architecture shown in Figure \ref{fig:similarity}. The model is trained as a binary classifier that receives two 30-second telemetry sequences $(900\times21)$ as input. Each of the sequences is first passed through an identical encoder using the LSTM funnel architecture described in \S\ref{sec:lstmfunnelarchitecture} to generate a 256-dimensional embedding. The Euclidean distance between these embeddings is then used to output a $1$ if the two motion sequences correspond to the same action, and a $0$ otherwise.

\vspace{-0.3em}
\vspace{-0.3em}

The approach illustrated in Figure \ref{fig:similarity} is sometimes known as a ``Siamese neural network'' \cite{Chicco2021}. Siamese architectures have previously been used in VR identification models \cite{miller_2021}, albeit with CNN layers rather than our LSTM funnel architecture.
An advantage of this approach is that while it is trained as a binary classifier for ``action similarity,'' a limb of the model can later be used on its own as an ``action encoder,'' such that the Euclidean distance between two embeddings produced by the encoder reveals the similarity of actions in the inputs.

To train the action similarity model, we randomly sampled 50,000 distinct pairs of ``similar'' motion sequences from the dataset of \S\ref{sec:dataset}, and another 50,000 distinct pairs of ``dissimilar'' motion sequences. An additional 5,000 similar and 5,000 dissimilar pairs were sampled for validation, with a further 5,000 similar and 5,000 dissimilar pairs for testing.
For the purpose of defining similarity, we use the ``software.activity.id'' attribute of the recordings provided in BOXRR-23 \cite{nair2023berkeley}. 
In this case, the attribute corresponds to the exact map the user is playing (see \S\ref{sec:beatsaber}).
In every instance, the two motion sequences constituting a pair of inputs originate from different users.
The model is thus tasked to classify whether two different users are playing identical or different in-game levels.

When training the action similarity model on the 200,000 motion sequences (50,000 pairs $\times$ 2 classes) discussed above, early stopping occurred after the 156th epoch.
The model achieved 100.00\% training accuracy, 99.53\% validation accuracy, and 99.40\% testing accuracy.
Therefore, we now have (1) a binary classifier that can determine with 99.4\% accuracy whether two motion sequences correspond to the same map, and (2) an action encoder that has learned an approximate metric for measuring the similarity of two motion sequences.

\eject

\subsection{User Similarity}
\label{sec:usersimilarity}

Next, we train a ``user similarity'' model, which is essentially the inverse of the action similarity model described above. Using the same architecture as before (Figure \ref{fig:similarity}), we now 
randomly sampled 50,000 pairs of motion sequences from the same user, and another 50,000 distinct pairs of motion sequences from different users. Again, an additional 5,000 similar and 5,000 dissimilar pairs were sampled for validation, and 5,000 similar and 5,000 dissimilar pairs for testing.
In every instance, the two motion sequences constituting a pair of inputs originate from different in-game maps.
The model is thus now tasked to ignore the action and classify whether two motion samples originate from the same or different users.

When training the user similarity model on the 200,000 motion sequences (50,000 pairs $\times$ 2 classes) discussed above, early stopping occurred after the 27th epoch.
The model achieved 97.94\% training accuracy, 92.60\% validation accuracy, and 92.81\% testing accuracy.
Therefore, in addition to the (1) action similarity and (2) action encoder models, we also have (3) a user similarity classifier that can determine with 92.8\% accuracy whether two motion sequences correspond to the same user, and (4) a user encoder that has learned a metric for characterizing the user from a motion sequence.

\subsection{Anonymizer}
\label{sec:anonymizer}

\begin{figure}[h]
    \centering
    \includegraphics[width=\linewidth]{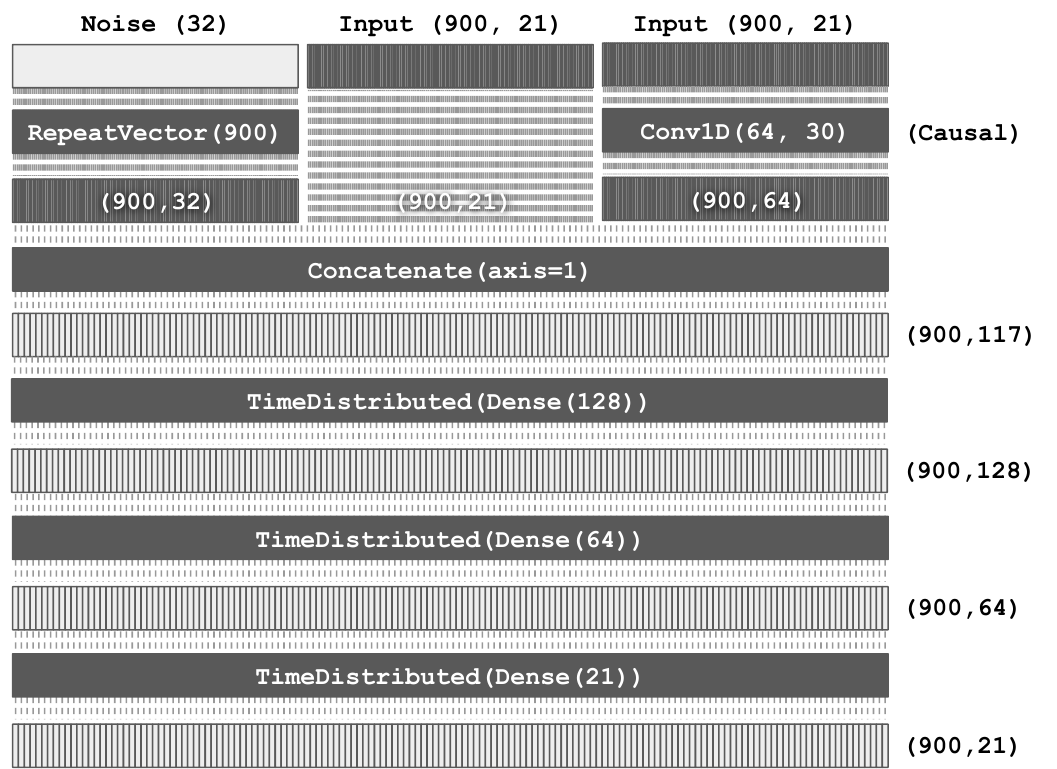}
    \caption{Architecture used for anonymizer model.}
    \label{fig:anonymizer}
\end{figure}

Using the trained action similarity and user similarity models described above, we can now train the ``anonymizer'' model that performs the core deep motion masking functionality.
The anonymizer model receives as input a 30-second motion telemetry sequence $(900\times21)$, and a 32-dimensional noise vector containing random Gaussian noise. It uses these values to output a corresponding 30-second motion sequence $(900\times21)$ that is an anonymized version of the input. Our anonymizer model architecture is illustrated in Figure \ref{fig:anonymizer}.

\eject

In addition to the motion input $(900\times21)$ and noise $(32)$ (which is repeated to produce a $(900\times32)$ sequence), a learned 1D convolution $(900\times64)$ of the motion input is produced. 
These three sequences are then vertically concatenated to produce a $(900\times117)$ hybrid sequence. Multiple time-distributed dense layers are then used to reduce this sequence back to a $(900\times21)$ output motion sequence.

The intuition behind this architecture is that the dense layers effectively combine the noise and motion data to anonymize the motion data in a way that is consistent across each frame, creating a smooth and continuous motion output. This allows the motion to be anonymized in 3D space, but not across the time domain. Therefore, the 1D convolution is added to allow limited manipulation of time-series relationships in the data within a sliding one-second window.

Importantly, every component of this architecture respects causality; the model does not have the capability to ``look into the future'' when producing any output frame. For example, the 1D convolution uses causal padding such that only frames $N-30$ through $N$ are used in the output of frame $N$.
After training, this allows the resulting anonymizer model to be deployed in real-time on a frame-by-frame basis.

\begin{figure}[h]
    \centering
    \includegraphics[width=\linewidth]{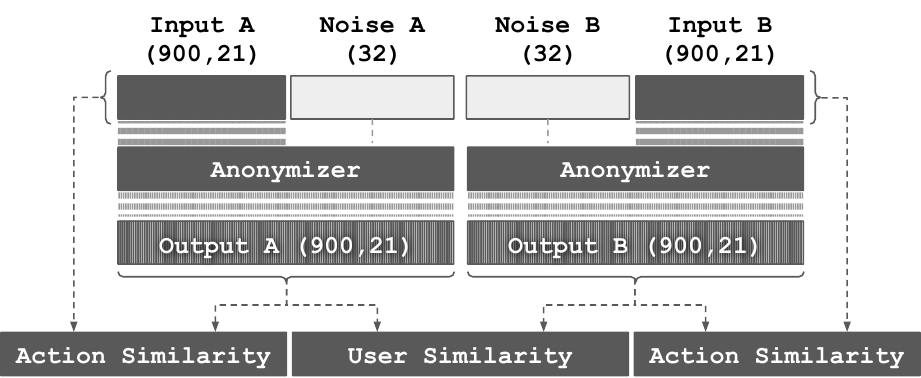}
    \caption{Siamese architecture for training anonymizer model.}
    \label{fig:anonymizer2}
\end{figure}

Figure \ref{fig:anonymizer2} shows how the action similarity and user similarity models are used to train the anonymizer model. First, the anonymizer is pre-trained for 20 epochs as an autoencoder with MSE loss, such that the output frames are initially nearly identical to the inputs, regardless of which noise values are provided. Then, a Siamese architecture is once again used. Leveraging the trained action and user similarity models (the weights of which are now frozen), the anonymizer is trained with the following loss function components:

\begin{enumerate}[leftmargin=*]
    \itemsep 0em
    \item The action embedding of $\mathsf{input}_A$ and $\mathsf{output}_A$ should always be as close as possible (irrespective of $\mathsf{noise}_A$).
    \item Similarly, $\mathsf{input}_B$ and $\mathsf{output}_B$ should always have as close of an action embedding as possible.
    \item If $\mathsf{user}_A = \mathsf{user}_B$ and $\mathsf{noise}_A = \mathsf{noise}_B$, the user embedding for $\mathsf{output}_A$ and $\mathsf{output}_B$ should be as close as possible.
    \item If $\mathsf{user}_A = \mathsf{user}_B$ and $\mathsf{noise}_A \neq \mathsf{noise}_B$, the user embedding for $\mathsf{output}_A$ and $\mathsf{output}_B$ should be far apart.
\end{enumerate}

\eject

In other words, the action represented by an anonymized motion sequence should remain unchanged from the original motion sequence, helping to achieve the indistinguishability goal of our model.
Furthermore, the intended use of the noise value is to be randomly sampled at the start of each new session, and then to remain consistent within that session. Thus, a user should assume a consistent faux identity within a session, but should assume distinct apparent identities across sessions, achieving cross-session unlinkability.
Importantly, by using the adversarial training method in Figure \ref{fig:anonymizer2}, the anonymizer receives precise differentiable feedback from the action and user similarity models on how to achieve both of these goals.

An additional advantage of this training method is that it provides a tunable security parameter that can be used to adjust the balance of anonymity and usability while training the model. If additional usability is needed, more weight can be placed on loss components (1) and (2), causing the output motion to appear more similar to the input motion. On the other hand, if more anonymity is required, further weight can be put on loss components (3) and (4), emphasizing cross-session unlinkability of outputs. In our evaluation, we use equal weights for both components, meaning that indistinguishability and cross-session unlinkability are equally important goals.

To train the anonymizer, we randomly sampled 50,000 pairs of motion sequences, with both samples in any given pair coming from the same user. We then randomly sampled 50,000 pairs of random Gaussian noise vectors. For half of the pairs, the noise inputs are identical ($\mathsf{noise}_A = \mathsf{noise}_B$), while for the other half, they are different ($\mathsf{noise}_A \neq \mathsf{noise}_B$), per the loss function described above. 
An additional 5,000 pairs were sampled for testing. No validation set was used; the model was trained for a full 500 epochs without early stopping.

The model achieved user similarity accuracy of 95.54\% on the training data and 94.71\% on the testing data. 
In other words, 94.71\% of the time, the model correctly predicted that $\mathsf{user}_A = \mathsf{user}_B$ when $\mathsf{noise}_A = \mathsf{noise}_B$ and that $\mathsf{user}_A \neq \mathsf{user}_B$ when $\mathsf{noise}_A \neq \mathsf{noise}_B$.
These numbers should be interpreted in light of the user similarity model's baseline accuracy of 92.81\%.
Importantly, on both datasets, the model achieved an action similarity accuracy of 100.00\%; in every training and testing sample, the action similarity model correctly described the input and output motion as containing the same action.

\subsection{Normalizer}
\label{sec:normalizer}

\begin{figure}[h]
    \centering
    \includegraphics[width=\linewidth]{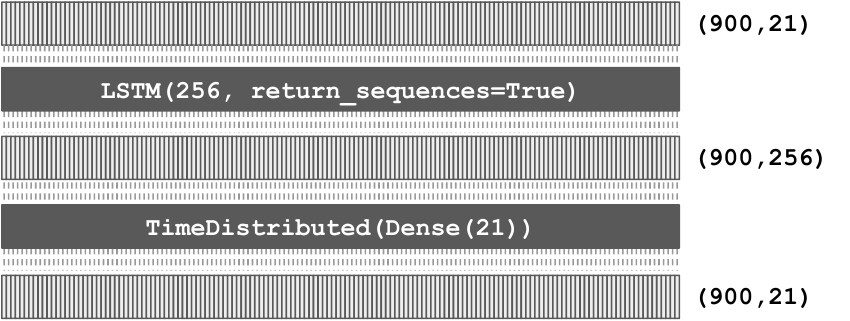}
    \caption{Normalizer model architecture.}
    \label{fig:normalizer}
\end{figure}

While the anonymizer is effective at obscuring the identity of a VR user while keeping their big-picture actions looking the same, it introduces some undesirable noise to the telemetry signal (at the frame level) due to the lack of an incentive against doing so.
One idea for combating this would be to use an adversarial architecture (e.g., GAN \cite{goodfellow2020generative}) with a discriminator network that provides feedback to the anonymizer by attempting to distinguish anonymized motion from unmodified motion sequences. 
Unfortunately, we found this idea difficult to apply for our use case as discussed further in \S\ref{sec:futurework}. Instead, we use a normalizer model that aims to reverse the effects of the anonymizer using the architecture in Figure \ref{fig:normalizer}.

The normalizer receives as input an anonymized motion sequence $(900\times21)$ and outputs a normalized motion sequence $(900\times21)$. The relatively simple architecture consists of an LSTM layer that returns a 256-dimensional state for each frame and a time-distributed dense layer that converts each state back to a 21-dimensional output. As with the anonymizer, the architecture obeys causality (e.g., no bidirectional layers) and can therefore be deployed in a real-time setting.

To train the normalizer, we randomly sampled 50,000 motion sequences from random users and maps and anonymized each of them using random noise vectors. We then trained the normalizer using a subset of the anonymized motion sequences as inputs and the corresponding original motion sequences as the target outputs, with a mean squared error loss function. Using a portion of the sequences reserved for testing, we found that the mean squared error between input and output samples after z-score normalizing every dimension was reduced by about one order of magnitude.

Importantly, the normalizer model is not provided with the noise values used to anonymize the original motion sequences, and, during inference, does not have access to the original motion data. Therefore, it will never be able to fully recover the original motion sequences, and cannot reduce the anonymity of the motion sequences, as any deterministic algorithm that could undo the anonymization without access to the original motion or noise values could also be deployed by an adversary to defeat anonymized motion sequences. 
Instead, the normalizer network can only remove any component of the noise added by the anonymizer
that is consistent or predictable across all anonymized motion sequences, which does not affect the actions or anonymity of any particular user.

The entire deep motion masking system architecture, with about 2.2 million parameters, is shown in \S\ref{app:architecture}. Of these, 290k parameters are in the normalizer, with the action and user similarity models containing nearly one million parameters each. The anonymizer itself contains only about 65k trainable parameters, allowing it to run extremely quickly on its own.

\eject
\subsection{Deployment}
\label{sec:deployment}
\vspace{-0.5em}

Deploying the trained models for post-hoc anonymization of motion recordings is now as simple as randomly sampling 32 Gaussian noise values, invoking the anonymizer model on the input motion sequence and noise values, and then running the normalizer model on the output of the anonymizer.

Based on our observations, we suggest a few simple optimizations to the above process.
First, we observe that it is better for indistinguishability if the population mean and standard deviation of each motion dimension in anonymized recordings match the 
population mean and standard deviation of each motion dimension in unmodified motion.
This population-level shift does not impact the anonymity of any individual user. Second, we recommend duplicating the first frame of motion 30 times before including the subsequent motion input. This ensures the 1D convolution buffer of the anonymizer model is always filled with real data,
reducing apparent noise and instability in the first second of the anonymized output. Finally, the quaternions representing rotational dimensions of the output should be normalized to unit magnitude to maintain validity.

The deep motion masking system can also be used in a real-time (streaming) setting. To do so, a buffer of the last 30 frames should be maintained and initially filled with 30 copies of the first frame. For each new frame, a corresponding anonymized frame can be produced by running the anonymizer's learned 1D convolution on the frame buffer, then concatenating its 64-dimensional output to the 21-dimensional input and 32-dimensional noise vector to produce a 117-dimensional hybrid vector. 
That hybrid vector can then be converted into a 21-dimensional anonymized output frame using the dense layers of the anonymizer.

Next, the optional optimization of shifting the population mean and standard deviation of each motion dimension back to that of the general population can be applied. Finally, the resulting frame can be fed into the LSTM layer of the normalizer, and the 256-dimensional LSTM state can be used by the dense layer of the normalizer to recover a final 21-dimensional anonymized and normalized output frame. Again, the quaternions should be normalized to unit magnitude.

Overall, the real-time deployment of deep motion masking adds no delay other than the computational delay of invoking the anonymizer and normalizer models, which we found to be less than 1~ms.
Due to the causal design of the architecture, the anonymized and normalized output in the streaming setting is identical to the result of the post-hoc anonymization process.
\vspace{-1.5em}
\section{Evaluation}
\label{sec:evaluation}
\vspace{-0.5em}

Having fully described our proposed deep motion masking approach, we now present a detailed evaluation of the privacy and usability of the resulting system. Our evaluation directly compares the cross-session unlinkability and indistinguishability of our system to that of MetaGuard \cite{nair2023going}, the prior state-of-the-art system for anonymizing VR motion data.

\eject

\subsection{Anonymity}
\label{sec:anonymity}

First, we analyze the impact of our deep motion masking system on cross-session linkability. If the system is effective at anonymizing VR motion data, it should be able to trick our LSTM funnel classification model (\S\ref{sec:lstmfunnelarchitecture}) into wrongly classifying anonymized users in most instances.
However, to ensure that our anonymizer didn't overfit by only fooling our own classification model, we also include the Random Forest identification model of Miller et al. \cite{miller_personal_2020} and LightGBM-based identification model of Nair et al. \cite{291259}.

Furthermore, we train each model both as an oblivious adversary, which is trained on unmodified motion sequences from each user and tested on anonymized motion sequences, and as an adaptive adversary, which is trained on anonymized motion sequences from within a session and tested from anonymized motion sequences in another session.
Per our definition of cross-session unlinkability in \S\ref{sec:problemstatement}, none of the models are trained on multiple independent sessions of anonymized motion, as we operate under the assumption that no external identifiers can be used to link sessions together.

To perform the evaluation, we randomly selected 1,000 users from the dataset of \S\ref{sec:dataset}. 
In order to be representative of average VR users, we only include users for which between 30 and 100 recordings were present; about 20,000 such users exist in the dataset. For each user, we selected 10 recordings to constitute the first session (for training) and another 10 recordings to constitute the second session (for testing). 
We then anonymized either one or both sessions (depending on the type of adversary), using either MetaGuard or the full post-hoc anonymization pipeline detailed in \S\ref{sec:deployment}. The results of training and testing each of the considered identification models on each set of data are summarized in Table \ref{tab:anonymity} below.

\begin{table}[h]
\centering
\resizebox{\columnwidth}{!}{%
\begin{tabular}{l|ll|ll|ll|}
\cline{2-7}
 & \multicolumn{2}{c|}{\textbf{Miller et al. \cite{miller_2020}}} & \multicolumn{2}{c|}{\textbf{Nair et al. \cite{291259}}} & \multicolumn{2}{c|}{\textbf{LSTM Funnel (\S\ref{sec:lstmfunnelarchitecture})}} \\ \cline{2-7} 
 & \multicolumn{1}{l|}{\textit{Oblivious}} & \textit{Adaptive} & \multicolumn{1}{l|}{\textit{Oblivious}} & \textit{Adaptive} & \multicolumn{1}{l|}{\textit{Oblivious}} & \textit{Adaptive} \\ \hline
\multicolumn{1}{|l|}{Unmodified} & \multicolumn{1}{l|}{90.3\%} & 90.3\% & \multicolumn{1}{l|}{91.0\%} & 91.0\% & \multicolumn{1}{l|}{96.5\%} & 96.5\% \\ \hline
\multicolumn{1}{|l|}{MetaGuard \cite{nair2023going}} & \multicolumn{1}{l|}{57.4\%} & 79.5\% & \multicolumn{1}{l|}{67.0\%} & 84.3\% & \multicolumn{1}{l|}{81.3\%} & 96.3\% \\ \hline
\multicolumn{1}{|l|}{DMM (\S\ref{sec:method})} & \multicolumn{1}{l|}{1.5\%} & 1.2\% & \multicolumn{1}{l|}{3.1\%} & 3.5\% & \multicolumn{1}{l|}{3.7\%} & 0.1\% \\ \hline
\end{tabular}%
}
\caption{Identification accuracy for various adversaries and model architectures, with and without anonymization.}
\label{tab:anonymity}
\end{table}

As demonstrated by the results of Table \ref{tab:anonymity}, deep motion masking is significantly better than MetaGuard at anonymizing users across sessions. 
While MetaGuard users remain up to 96\% identifiable, deep motion masking reduces identification accuracy to less than 4\%, representing a $20\times$ to over $100\times$ improvement in anonymity depending on the model.

As expected, adaptive adversaries are usually better at identifying anonymized users across sessions, as information about what the user looks like when using the anonymity tool of choice (albeit with different noise values) can be incorporated into the identification model. In the case of MetaGuard, this allows the LSTM funnel architecture to perform at nearly full accuracy, as the model learns to ignore anonymized dimensions and identify users by the unmodified dimensions.

\eject

Interestingly, however, the LSTM funnel model actually performs significantly worse with the deep motion masking samples when trained adaptively.
This is likely because component (3) of the loss function used to train the anonymizer model (\S\ref{sec:anonymizer}) is measured by a user encoder based on the LSTM funnel architecture. The anonymizer model therefore is particularly good at tricking the LSTM funnel architecture into learning fictitious user attributes and consequently becoming worse at identifying users.

\subsection{Usability}
\label{sec:usability}

Next, to evaluate the indistinguishability of motion data anonymized with deep motion masking, we conducted a large-user study (N=182).
The study consisted of an online survey in which users were asked to watch VR motion recordings from the game Beat Saber in the Beat Saber web replay viewer tool \cite{viewer} after reading and agreeing to an informed consent document. Four types of treatments were tested:

\begin{enumerate}[leftmargin=*]
    \itemsep 0em
    \item As a negative control group, we included unmodified VR motion recordings from the dataset of \S\ref{sec:dataset} that will certainly be indistinguishable from natural human motion.
    \item As a positive control group, we included completely AI-generated motion recordings created by CyberRamen \cite{cyberramen}, a machine learning model trained to play Beat Saber. As it stands, these recordings are easily distinguishable from natural motion, serving as a good test of response quality.
    \item As a baseline treatment group, we included recordings anonymized using MetaGuard \cite{nair2023going} with the ``height,'' ``wingspan,'' and ``room size'' defenses enabled at the ``medium'' privacy settings suggested by the authors.
    \item As our new treatment group, we included recordings anonymized with deep motion masking using the same models and processes as the anonymity evaluation (\S\ref{sec:anonymity}).
\end{enumerate}

\begin{figure}[h]
    \centering
    \includegraphics[width=\linewidth]{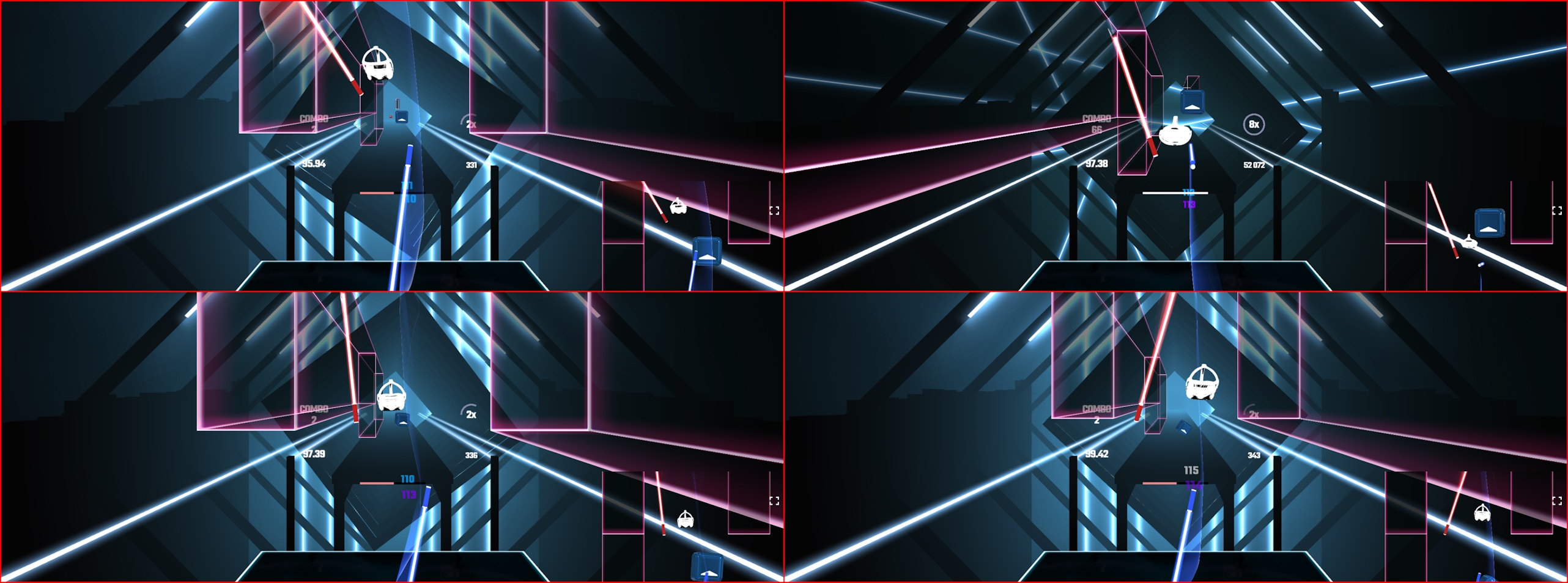}
    \caption{A set of Beat Saber replays shown to participants.}
    \label{fig:replays}
\end{figure}

As shown in Figure \ref{fig:replays}, users were given one set of recordings at a time, consisting of four recordings of different users playing the same map in Beat Saber (see \S\ref{sec:beatsaber}).
To remove confounding variables, all recordings in all sets were first normalized to 30 FPS and trimmed to the first 30 seconds.
One of the four recordings in each set was additionally treated (i.e., ``anonymized'') using one of the four treatments listed above.

Each user was shown 12 such sets of recordings in a randomized order, corresponding to a slow, medium, and fast song for each of the four treatment groups described above. For each set, their task was to decide which (if any) of the four recordings was modified.
To aid their decision, users could view each replay in slow motion,
zoom in on particular areas,
and turn to view the motion from a variety of perspectives.

When recruiting participants for our study, we focused primarily on finding VR users with significant Beat Saber experience, as such users are more likely to be familiar with what natural VR motion data should look like, and thus are likely to be more challenging and discerning critics of our system. With that in mind, we primarily recruited participants through social media pages related to VR, 
and through VR interest groups like CVRE \cite{cvre}. However, we also wanted to ensure that some number of novice users participated in the study, 
and recruited a small number of participants from a broader general population for that purpose.

\begin{figure}[h]
    \centering
    \includegraphics[width=\linewidth]{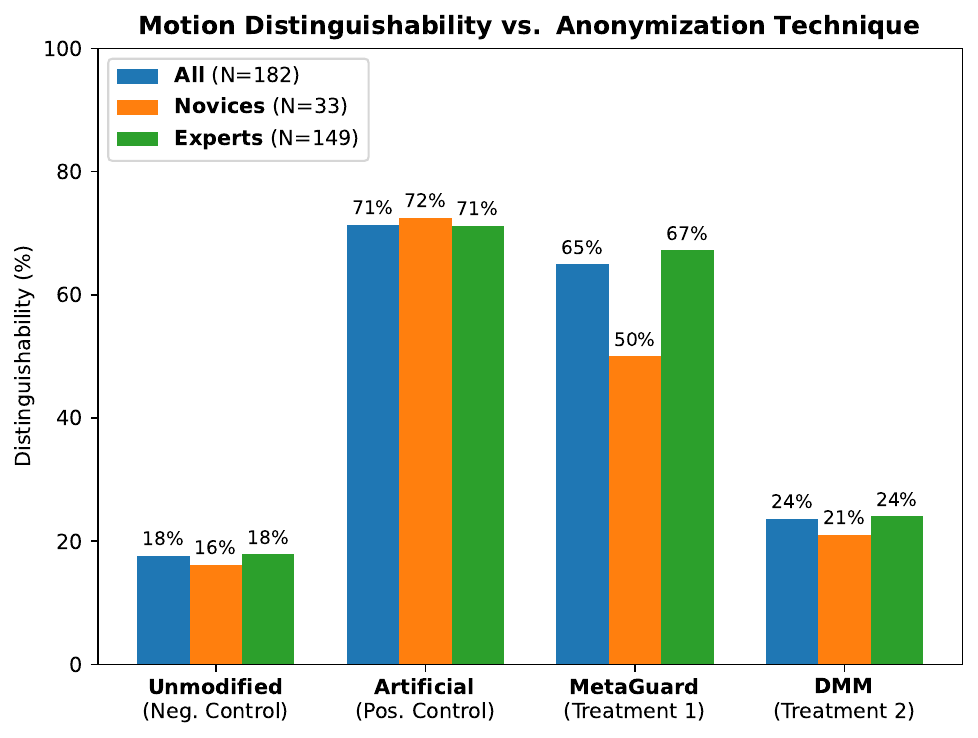}
    \caption{Results of indistinguishability user study.}
    \label{fig:results}
\end{figure}

The study ran for two weeks, from September 20th, 2023 through October 3rd, 2023, and received 241 responses in that time.
We removed the 59 responses that were either blank or answered all six of the control questions incorrectly, leaving 182 valid responses.
Of those, 149 were from expert Beat Saber players (with 100 or more hours of in-game experience), and the remaining 33 participants were novices (with 0 to 100 hours of experience). Figure \ref{fig:results} shows the observed distinguishability for each of the evaluated treatments.

The negative control group has a surprisingly high rate of distinguishability in our results (18\%). This indicates that when unsure about which replay was modified, users in our study were prone to randomly guessing one of the four replays rather than indicating that all four replays were unmodified.

\eject

With that in mind, unfortunately, the replays anonymized with deep motion masking were still not perfectly indistinguishable from natural motion, but were only marginally more distinguishable than the negative control group. Moreover, deep motion masking represents a significant improvement over the MetaGuard \cite{nair2023going} system, with nearly a $3\times$ reduction in the rate of distinguishability, particularly for expert users. Using both a standard ${\chi}^2$ test and Fisher's exact test \cite{fisher}, the difference between MetaGuard and deep motion masking is highly statistically significant with $p < 0.01$.

\subsection{Interactivity}
\label{sec:interactivity}

The indistinguishability study of \S\ref{sec:usability} already demonstrates that the deep motion masking anonymizer has minimal impact on observed interactions between users and virtual objects, as participants in that study could view users interacting with virtual objects (namely, blocks in the Beat Saber game) when determining whether a motion sample was modified.
However, to enhance the explainability of the user study results and further demonstrate that our deep motion masking system satisfies the stated goal of interactivity, we conducted additional in vitro experiments in which we simulated the effects of deep motion masking on interactions with virtual objects.

\begin{figure}[h]
    \centering
    \includegraphics[width=\linewidth]{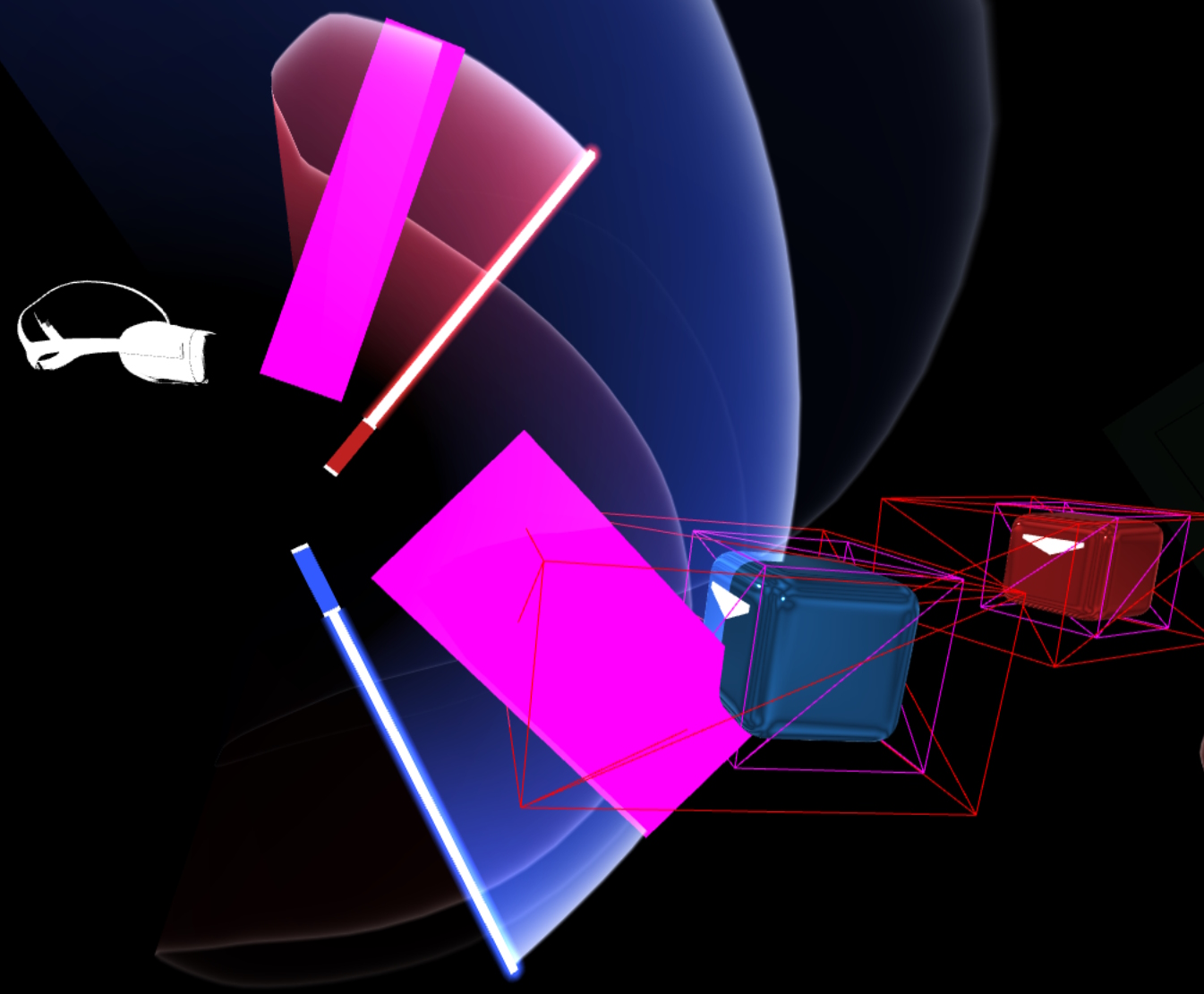}
    \caption{Collision modeling for Beat Saber objects.}
    \label{fig:simsaber}
\end{figure}

SimSaber \cite{simsaber} is a Python library that simulates Beat Saber gameplay by faithfully replicating the physics and collision detection algorithms used by Beat Saber and the Unity game engine \cite{unity}, as shown in Figure \ref{fig:simsaber}.
We randomly sampled 1,000 Beat Saber replays from the dataset of \S\ref{sec:dataset}, and anonymized them with deep motion masking. We then ran the original and modified replays through the simulator to evaluate what impact the anonymization process had on user interactions with the virtual blocks in Beat Saber.

\eject

In Beat Saber, the cutting of a block with a saber is typically characterized by the player's pre-swing angle, post-swing angle, and accuracy (closeness to the center of the block). The combination of these three factors is used to calculate the player's score. At a minimum, a usable anonymization tool should not significantly impact these three measurements in order to avoid substantially affecting the user's performance.

In our evaluation of 1,000 replays, we found that anonymized players had a mean absolute difference in pre-swing angle of about $5^{\circ}$, and an average relative difference of 4.5\%. The mean absolute difference in post-swing angle was about $4^{\circ}$, and an average relative difference of 6.7\%.
The closeness to the center of the block was modified by a mean absolute difference of about $6.5$~cm, resulting in an average relative accuracy difference of 14\%.
Overall, the mean absolute difference in the player's score after anonymization was only 0.7\%, a difference that should be unnoticeable for all but the most experienced players. 
Still, our system may not be suitable for situations requiring extreme precision (see \S\ref{sec:discussion}).

These findings complement our indistinguishability results of \S\ref{sec:usability} by demonstrating that the deep motion masking system is able to maintain approximate apparent interactions with in-game objects, despite having no direct information about virtual object positions and geometries. By incorporating an action similarity metric (\S\ref{sec:actionsimilarity}), the model simply learns to avoid making changes that are likely to change the semantic meaning of the motion. As a result, viewers struggle to distinguish the anonymized motion from that of a real user.

\vspace{-0.8em}
\subsection{Ethics}
\label{sec:ethics}
\vspace{-0.6em}

The primary source of data for this study is the BOXRR-23 dataset \cite{nair2023berkeley}, a publicly available dataset intended for use in VR research, including security and privacy research. This dataset has already been used in published research papers in the VR security and privacy domain \cite{291259}.
It contains built-in privacy measures, such as pseudonymization of participants, and was reviewed by the legal and ethics boards of its authors prior to release.
We specifically only use the BeatLeader part of the dataset in our research; these users agree to the use of their data for ``research topics such as VR security, privacy, and usability'' in the BeatLeader privacy policy \cite{blpolicy}.

Other than the BOXRR-23 dataset, the only additional data used in this paper is from our usability study in \S\ref{sec:usability}. 
All participants in the survey were adults over the age of 18, and no vulnerable populations were specifically targeted in this study.
Participants consented to their inclusion in academic research by reading and agreeing to an informed consent document before proceeding in the survey.
Users optionally provided their Beat Saber username, but no further identifiable information was collected.
Information collected consisted exclusively of the users' selections of which recordings they believed were modified.
Therefore, the likelihood of any harm to participants, either through participating or through a later breach of confidentiality, is exceedingly low.

\eject

All aspects of this study, including our use of the public BeatLeader data and our collection of survey responses in \S\ref{sec:usability}, were also independently reviewed and approved by an OHRP-certified IRB under protocol number 2023-06-16467.
\section{Discussion}
\label{sec:discussion}
\vspace{-0.5em}

Anonymizing VR motion data inherently involves diverging from the original motion data to some extent. The approach detailed in \S\ref{sec:method} ensures that such deviations correspond mostly to apparent differences not in the actions being taken but rather in the user taking the actions.
This results in the system being highly suitable for motion data intended for consumption by human observers, as demonstrated in \S\ref{sec:usability}.

On the other hand, there will always be VR applications in which very high precision is required, 
such as telemedicine, competitive e-sports, or remote operation of equipment.
In such situations, the average discrepancies measured in \S\ref{sec:interactivity} of 6.5~cm (position) and $5^{\circ}$ (rotation) may be intolerable.
If anonymity is still desired in such an application, an alternative solution, such as secure multi-party computation or trusted execution environments, may be more suitable.
Thus, we recommend a two-channel approach for VR motion data, with one system handling real-time anonymization of low-fidelity motion for human eyes,
and another handling precise motion data for asynchronous computational use. Deep motion masking presents a secure, usable, and scalable solution for the former scenario, while the latter merits further investigation.

\subsection{Limitations}
\label{sec:limitations}
\vspace{-0.5em}

One major limitation of our system is that it has only been trained on data from a single VR application, Beat Saber.
This is because there are currently about four orders of magnitude more motion data available from Beat Saber than any other VR application, with deep learning models benefiting from large amounts of training data.
Unlike prior work using this dataset, we don't allow our model to see anything specific to Beat Saber, such as block positions and timings.
Therefore, it should be possible to train a deep motion masking model, using the present architecture, on motion data from any VR application, if enough motion data were available. However, without such data, we cannot confidently claim that the evaluation results will generalize to other applications.

Another major limitation of deep motion masking is that it loses the provable security properties of MetaGuard \cite{nair2023going}. One of the most significant features of MetaGuard is that it obeys $\varepsilon$-differential privacy, and thus provides provable security and privacy properties.
However, that provability only extends to the specific dimensions that the authors consider in the paper.
As demonstrated in \S\ref{sec:motivation}, this creates a weakness, as rotational dimensions are excluded entirely. Thus, while proving the security of our deep learning approach is significantly harder, the method empirically provides better cross-session unlinkability than MetaGuard as demonstrated in \S\ref{sec:anonymity}.

\eject

\subsection{Future Work}
\label{sec:futurework}

One important area of future work in this field is extending motion anonymization systems support to full-body tracking data. 
Deep motion masking is particularly suitable for this purpose, as it doesn't involve manually engineering features between pairs of tracked objects, and may in fact be immediately applicable to full-body telemetry streams.
At present, we lack a sufficiently large full-body motion capture dataset to use for training. However, as next-generation VR devices move towards full-body tracking, such data may become readily available, and the importance of full-body motion anonymization will simultaneously increase.

On the subject of data, future work may focus on procuring large-scale VR motion datasets from applications other than Beat Saber. Demonstrating the generalizability of deep motion masking to a wide variety of VR games and applications is an important step toward the potential adoption of such a system.

Finally, one may wonder why a discriminator network (i.e., a GAN \cite{goodfellow2020generative} architecture) was not used in this paper.
While GANs theoretically could be a great way to ensure anonymized motion data remains indistinguishable, we found them to not work well in practice for this dataset, because the goal of computational indistinguishability is too strong to be practical.
While data anonymized with our deep motion masking system is almost perfectly indistinguishable to the human eye (\S\ref{sec:usability}), it can still be distinguished by a machine learning classifier with almost 100\% accuracy.
Thus, regardless of which combinations of architectures and learning rates we tried, a GAN always resulted in the generator ceasing to make progress as the discriminator reached 100\% accuracy.

However, we leave open the possibility that a GAN could work in this application, and perhaps produce even better results, if used in a way that we did not consider.
Other architectures, such as diffusion or transformer models, could also be useful, although inference latency may become a concern.
We hope to see future work that explores various other architectures and techniques for masking VR motion data.
\section{Conclusion}
\label{sec:conclusion}

Deep learning is increasingly emerging as a powerful method for the usable real-time anonymization of sequential data (e.g., voice anonymization \cite{285349}).
In this paper, we've shown that deep learning can also be an effective tool for anonymizing VR telemetry data by developing a technique we call deep motion masking,
which is analogous to a real-time voice changer for movement patterns.
By decomposing the space of motion variability into action-related variation and user-related variation, our model is effective at hiding user identity while maintaining action similarity, leading to better indistinguishability and cross-session unlinkability than prior methods.

\eject

\clearpage

\section*{Acknowledgments}
We greatly appreciate the advice and support of Allen Yang, Ananya Kharche, Atticus Cull, Beni Issler, Bjoern Hartmann, Brandon Huang, Charles Dove, Chris He, Christian Rack, Dziugas Ramonas, Eric Paulos, James Smith, Rui Wang, Shuixian Li, Viktor Radulov, Xiaoyuan Liu, and Zade Lobo.
This work was supported in part by several organizations. The full list of organizations supporting this work will be included in any final publications.
Any opinions, findings, and conclusions or recommendations expressed in this material are those of the authors and do not necessarily reflect the views of their employers or the supporting entities. We sincerely thank all of the users who participated in our study or contributed to the BOXRR-23 dataset for making this work possible.

\section*{Availability}
The source code and documentation necessary to train and test all of the models and evaluations discussed in this paper are available on our GitHub repository under a BSD license:\\

\centerline{\url{https://github.com/metaguard/metaguardplus}}

\small
\bibliographystyle{plainurl}
\bibliography{references}

\begin{thebibliography}{10}

\bibitem{9319051}
Nadisha-Marie Aliman and Leon Kester.
\newblock Malicious design in aivr, falsehood and cybersecurity-oriented immersive defenses.
\newblock In {\em 2020 IEEE International Conference on Artificial Intelligence and Virtual Reality (AIVR)}, pages 130--137, 2020.
\newblock \href {https://doi.org/10.1109/AIVR50618.2020.00031} {\path{doi:10.1109/AIVR50618.2020.00031}}.

\bibitem{visionpro}
Apple {Vision} {Pro}.
\newblock URL: \url{https://www.apple.com/apple-vision-pro/}.

\bibitem{breiman2001random}
Leo Breiman.
\newblock Random forests.
\newblock {\em Machine learning}, 45:5--32, 2001.
\newblock \href {https://doi.org/10.1023/A:1010933404324} {\path{doi:10.1023/A:1010933404324}}.

\bibitem{priv_implications_VR}
Kent Bye, Diane Hosfelt, Sam Chase, Matt Miesnieks, and Taylor Beck.
\newblock The ethical and privacy implications of mixed reality.
\newblock In {\em ACM SIGGRAPH 2019 Panels}, SIGGRAPH '19, New York, NY, USA, 2019. Association for Computing Machinery.
\newblock \href {https://doi.org/10.1145/3306212.3328138} {\path{doi:10.1145/3306212.3328138}}.

\bibitem{Chicco2021}
Davide Chicco.
\newblock {\em Siamese Neural Networks: An Overview}, pages 73--94.
\newblock Springer US, New York, NY, 2021.
\newblock \href {https://doi.org/10.1007/978-1-0716-0826-5_3} {\path{doi:10.1007/978-1-0716-0826-5_3}}.

\bibitem{simsaber}
Atticus Cull and Vivek Nair.
\newblock {SimSaber}: {Python}-based {Beat} {Saber} replay simulator and scoring validator.
\newblock URL: \url{https://github.com/MetaGuard/SimSaber}.

\bibitem{cutting_recognizing_1977}
James~E. Cutting and Lynn~T. Kozlowski.
\newblock Recognizing friends by their walk: {Gait} perception without familiarity cues.
\newblock {\em Bulletin of the Psychonomic Society}, 9(5):353--356, May 1977.
\newblock \href {https://doi.org/10.3758/BF03337021} {\path{doi:10.3758/BF03337021}}.

\bibitem{cvre}
Collegiate {VR} {Esports} {League} ({CVRE}).
\newblock URL: \url{https://cvreleague.com/}.

\bibitem{de_guzman_security_2020}
Jaybie~A. De~Guzman, Kanchana Thilakarathna, and Aruna Seneviratne.
\newblock Security and privacy approaches in mixed reality: A literature survey.
\newblock \href {https://doi.org/10.1145/3359626} {\path{doi:10.1145/3359626}}.

\bibitem{285349}
Jiangyi Deng, Fei Teng, Yanjiao Chen, Xiaofu Chen, Zhaohui Wang, and Wenyuan Xu.
\newblock {V-Cloak}: Intelligibility-, naturalness- \& {Timbre-Preserving} {Real-Time} voice anonymization.
\newblock In {\em 32nd USENIX Security Symposium (USENIX Security 23)}, pages 5181--5198, Anaheim, CA, August 2023. USENIX Association.
\newblock URL: \url{https://www.usenix.org/conference/usenixsecurity23/presentation/deng-jiangyi-v-cloak}.

\bibitem{dwork_algorithmic_2013}
Cynthia Dwork and Aaron Roth.
\newblock The algorithmic foundations of differential privacy.
\newblock {\em Foundations and Trends in Theoretical Computer Science}, 9(3):211--407, 2013.
\newblock URL: \url{http://www.nowpublishers.com/articles/foundations-and-trends-in-theoretical-computer-science/TCS-042}, \href {https://doi.org/10.1561/0400000042} {\path{doi:10.1561/0400000042}}.

\bibitem{metaverse_privacy_1}
Ben Falchuk, Shoshana Loeb, and Ralph Neff.
\newblock The social metaverse: Battle for privacy.
\newblock {\em IEEE Technology and Society Magazine}, 37(2):52--61, 2018.
\newblock \href {https://doi.org/10.1109/MTS.2018.2826060} {\path{doi:10.1109/MTS.2018.2826060}}.

\bibitem{fisher}
R.~A. Fisher.
\newblock On the interpretation of x2 from contingency tables, and the calculation of p.
\newblock {\em Journal of the Royal Statistical Society}, 85(1):87--94, 1922.
\newblock URL: \url{http://www.jstor.org/stable/2340521}.

\bibitem{beat_saber}
Beat Games.
\newblock Beat {Saber}.
\newblock \url{https://beatsaber.com/}.
\newblock URL: \url{https://beatsaber.com/}.

\bibitem{garrido2024sok}
Gonzalo~Munilla Garrido, Vivek Nair, and Dawn Song.
\newblock Sok: Data privacy in virtual reality.
\newblock In {\em 24th Privacy Enhancing Technologies Symposium (PETS 24)}, 2024.

\bibitem{giaretta2022security}
Alberto Giaretta.
\newblock Security and privacy in virtual reality -- a literature survey, 2022.
\newblock \href {https://arxiv.org/abs/2205.00208} {\path{arXiv:2205.00208}}.

\bibitem{goodfellow2016deep}
Ian Goodfellow, Yoshua Bengio, and Aaron Courville.
\newblock {\em Deep learning}.
\newblock MIT press, 2016.
\newblock URL: \url{https://www.deeplearningbook.org/}.

\bibitem{goodfellow2020generative}
Ian Goodfellow, Jean Pouget-Abadie, Mehdi Mirza, Bing Xu, David Warde-Farley, Sherjil Ozair, Aaron Courville, and Yoshua Bengio.
\newblock Generative adversarial networks.
\newblock {\em Communications of the ACM}, 63(11):139--144, 2020.

\bibitem{hochreiter1997long}
Sepp Hochreiter and J{\"u}rgen Schmidhuber.
\newblock Long short-term memory.
\newblock {\em Neural computation}, 9(8):1735--1780, 1997.

\bibitem{boundedtruncatedIBM}
Naoise Holohan, Spiros Antonatos, Stefano Braghin, and Pól Mac~Aonghusa.
\newblock The {Bounded} {Laplace} {Mechanism} in {Differential} {Privacy}.
\newblock {\em Journal of Privacy and Confidentiality}, 10(1), December 2019.
\newblock \href {https://doi.org/10.29012/jpc.715} {\path{doi:10.29012/jpc.715}}.

\bibitem{jain_is_2016}
Eakta Jain, Lisa Anthony, Aishat Aloba, Amanda Castonguay, Isabella Cuba, Alex Shaw, and Julia Woodward.
\newblock Is the {Motion} of a {Child} {Perceivably} {Different} from the {Motion} of an {Adult}?
\newblock {\em ACM Transactions on Applied Perception}, 13(4):1--17, July 2016.
\newblock \href {https://doi.org/10.1145/2947616} {\path{doi:10.1145/2947616}}.

\bibitem{180378}
Suman Jana, David Molnar, Alexander Moshchuk, Alan Dunn, Benjamin Livshits, Helen~J. Wang, and Eyal Ofek.
\newblock Enabling {Fine-Grained} permissions for augmented reality applications with recognizers.
\newblock In {\em 22nd USENIX Security Symposium (USENIX Security 13)}, pages 415--430, Washington, D.C., August 2013. USENIX Association.
\newblock URL: \url{https://www.usenix.org/conference/usenixsecurity13/technical-sessions/presentation/jana}.

\bibitem{ke_lightgbm_2017}
Guolin Ke, Qi~Meng, Thomas Finley, Taifeng Wang, Wei Chen, Weidong Ma, Qiwei Ye, and Tie-Yan Liu.
\newblock {LightGBM}: {A} {Highly} {Efficient} {Gradient} {Boosting} {Decision} {Tree}.
\newblock In I.~Guyon, U.~Von Luxburg, S.~Bengio, H.~Wallach, R.~Fergus, S.~Vishwanathan, and R.~Garnett, editors, {\em Advances in {Neural} {Information} {Processing} {Systems}}, volume~30. Curran Associates, Inc., 2017.
\newblock URL: \url{https://proceedings.neurips.cc/paper/2017/file/6449f44a102fde848669bdd9eb6b76fa-Paper.pdf}.

\bibitem{keras}
Keras: Deep learning for humans.
\newblock URL: \url{https://keras.io/}.

\bibitem{291287}
Yoonsang Kim, Sanket Goutam, Amir Rahmati, and Arie Kaufman.
\newblock Erebus: Access control for augmented reality systems.
\newblock In {\em 32nd USENIX Security Symposium (USENIX Security 23)}, pages 929--946, Anaheim, CA, August 2023. USENIX Association.
\newblock URL: \url{https://www.usenix.org/conference/usenixsecurity23/presentation/kim-yoonsang}.

\bibitem{kingma2014adam}
Diederik~P Kingma and Jimmy Ba.
\newblock Adam: A method for stochastic optimization.
\newblock {\em arXiv preprint arXiv:1412.6980}, 2014.

\bibitem{kozlowski_recognizing_1977}
Lynn~T. Kozlowski and James~E. Cutting.
\newblock Recognizing the sex of a walker from a dynamic point-light display.
\newblock {\em Perception \& Psychophysics}, 21(6):575--580, November 1977.
\newblock \href {https://doi.org/10.3758/BF03198740} {\path{doi:10.3758/BF03198740}}.

\bibitem{kupin}
Alex Kupin, Benjamin Moeller, Yijun Jiang, Natasha Banerjee, and Sean Banerjee.
\newblock {\em Task-Driven Biometric Authentication of Users in Virtual Reality (VR) Environments: 25th International Conference, MMM 2019, Thessaloniki, Greece, January 8–11, 2019, Proceedings, Part I}, pages 55--67.
\newblock Springer, 01 2019.
\newblock \href {https://doi.org/10.1007/978-3-030-05710-7_5} {\path{doi:10.1007/978-3-030-05710-7_5}}.

\bibitem{10.1145/333329.333344}
Joseph~J. LaViola.
\newblock A discussion of cybersickness in virtual environments.
\newblock {\em SIGCHI Bull.}, 32(1):47–56, jan 2000.
\newblock \href {https://doi.org/10.1145/333329.333344} {\path{doi:10.1145/333329.333344}}.

\bibitem{priv_metaverse}
Ronald Leenes.
\newblock Privacy in the metaverse.
\newblock In Simone Fischer-H{\"u}bner, Penny Duquenoy, Albin Zuccato, and Leonardo Martucci, editors, {\em The Future of Identity in the Information Society}, pages 95--112, Boston, MA, 2008. Springer US.

\bibitem{liebers23}
Jonathan Liebers.
\newblock Exploring the stability of behavioral biometrics in virtual reality in a remote field study: Towards implicit and continuous user identification through body movements.
\newblock 10 2023.
\newblock \href {https://doi.org/10.1145/3611659.3615696} {\path{doi:10.1145/3611659.3615696}}.

\bibitem{10.1145/3411764.3445528}
Jonathan Liebers, Mark Abdelaziz, Lukas Mecke, Alia Saad, Jonas Auda, Uwe Gruenefeld, Florian Alt, and Stefan Schneegass.
\newblock Understanding user identification in virtual reality through behavioral biometrics and the effect of body normalization.
\newblock In {\em Proceedings of the 2021 CHI Conference on Human Factors in Computing Systems}, CHI '21, New York, NY, USA, 2021. Association for Computing Machinery.
\newblock \href {https://doi.org/10.1145/3411764.3445528} {\path{doi:10.1145/3411764.3445528}}.

\bibitem{liebers2022}
Jonathan Liebers, Sascha Brockel, Uwe Gruenefeld, and Stefan Schneegaß.
\newblock Identifying users by their hand tracking data in augmented and virtual reality.
\newblock {\em International Journal of Human-Computer Interaction}, pages 1--16, 10 2022.
\newblock \href {https://doi.org/10.1080/10447318.2022.2120845} {\path{doi:10.1080/10447318.2022.2120845}}.

\bibitem{quest3}
Meta {Quest} 3: {New} {Mixed} {Reality} {VR} {Headset}.
\newblock URL: \url{https://www.meta.com/quest/quest-3/}.

\bibitem{miller2023largescale}
Mark~Roman Miller, Eugy Han, Cyan DeVeaux, Eliot Jones, Ryan Chen, and Jeremy~N. Bailenson.
\newblock A large-scale study of personal identifiability of virtual reality motion over time, 2023.
\newblock \href {https://arxiv.org/abs/2303.01430} {\path{arXiv:2303.01430}}.

\bibitem{miller_personal_2020}
Mark~Roman Miller, Fernanda Herrera, Hanseul Jun, James~A. Landay, and Jeremy~N. Bailenson.
\newblock Personal identifiability of user tracking data during observation of 360-degree {VR} video.
\newblock {\em Scientific Reports}, 10(1):17404, October 2020.
\newblock Number: 1 Publisher: Nature Publishing Group.
\newblock \href {https://doi.org/10.1038/s41598-020-74486-y} {\path{doi:10.1038/s41598-020-74486-y}}.

\bibitem{miller_2020}
Robert Miller, Natasha Banerjee, and Sean Banerjee.
\newblock Within-system and cross-system behavior-based biometric authentication in virtual reality.
\newblock In {\em 2020 IEEE Conference on Virtual Reality and 3D User Interfaces Abstracts and Workshops (VRW)}, pages 311--316, 03 2020.
\newblock \href {https://doi.org/10.1109/VRW50115.2020.00070} {\path{doi:10.1109/VRW50115.2020.00070}}.

\bibitem{miller_2021}
Robert Miller, Natasha~Kholgade Banerjee, and Sean Banerjee.
\newblock Using siamese neural networks to perform cross-system behavioral authentication in virtual reality.
\newblock In {\em 2021 IEEE Virtual Reality and 3D User Interfaces (VR)}, pages 140--149, 2021.
\newblock \href {https://doi.org/10.1109/VR50410.2021.00035} {\path{doi:10.1109/VR50410.2021.00035}}.

\bibitem{9583839}
Alec~G. Moore, Ryan~P. McMahan, Hailiang Dong, and Nicholas Ruozzi.
\newblock Personal identifiability and obfuscation of user tracking data from {VR} training sessions.
\newblock In {\em 2021 IEEE International Symposium on Mixed and Augmented Reality (ISMAR)}, pages 221--228, 2021.
\newblock \href {https://doi.org/10.1109/ISMAR52148.2021.00037} {\path{doi:10.1109/ISMAR52148.2021.00037}}.

\bibitem{nair2023going}
Vivek Nair, Gonzalo~Munilla Garrido, and Dawn Song.
\newblock Going incognito in the metaverse: Achieving theoretically optimal privacy-usability tradeoffs in {VR}.
\newblock In {\em 36th ACM Symposium on User Interface Software and Technology (UIST 23)}, 2023.

\bibitem{nair_exploring_2023}
Vivek Nair, Gonzalo~Munilla Garrido, Dawn Song, and James O'Brien.
\newblock Exploring the {Privacy} {Risks} of {Adversarial} {VR} {Game} {Design}.
\newblock In {\em 23rd Privacy Enhancing Technologies Symposium (PETS 23)}, 2023.
\newblock \href {https://doi.org/10.56553/popets-2023-0108} {\path{doi:10.56553/popets-2023-0108}}.

\bibitem{291259}
Vivek Nair, Wenbo Guo, Justus Mattern, Rui Wang, James~F. O{\textquoteright}Brien, Louis Rosenberg, and Dawn Song.
\newblock Unique identification of 50,000+ virtual reality users from head \& hand motion data.
\newblock In {\em 32nd USENIX Security Symposium (USENIX Security 23)}, pages 895--910, Anaheim, CA, August 2023. USENIX Association.
\newblock URL: \url{https://www.usenix.org/conference/usenixsecurity23/presentation/nair-identification}.

\bibitem{nair2023berkeley}
Vivek Nair, Wenbo Guo, Rui Wang, James~F. O'Brien, Louis Rosenberg, and Dawn Song.
\newblock Berkeley open extended reality recordings 2023 ({BOXRR}-23): 4.7 million motion capture recordings from 105,852 extended reality device users, 2023.
\newblock \href {https://arxiv.org/abs/2310.00430} {\path{arXiv:2310.00430}}.

\bibitem{nair2023inferring}
Vivek Nair, Christian Rack, Wenbo Guo, Rui Wang, Shuixian Li, Brandon Huang, Atticus Cull, James~F. O'Brien, Marc Latoschik, Louis Rosenberg, and Dawn Song.
\newblock Inferring private personal attributes of virtual reality users from head and hand motion data, 2023.
\newblock \href {https://arxiv.org/abs/2305.19198} {\path{arXiv:2305.19198}}.

\bibitem{nair2023truth}
Vivek Nair, Louis Rosenberg, James~F. O'Brien, and Dawn Song.
\newblock Truth in motion: The unprecedented risks and opportunities of extended reality motion data, 2023.
\newblock \href {https://arxiv.org/abs/2306.06459} {\path{arXiv:2306.06459}}.

\bibitem{pfeuffer}
Ken Pfeuffer, Matthias~J. Geiger, Sarah Prange, Lukas Mecke, Daniel Buschek, and Florian Alt.
\newblock Behavioural biometrics in vr: Identifying people from body motion and relations in virtual reality.
\newblock In {\em Proceedings of the 2019 CHI Conference on Human Factors in Computing Systems}, CHI '19, page 1–12, New York, NY, USA, 2019. Association for Computing Machinery.
\newblock \href {https://doi.org/10.1145/3290605.3300340} {\path{doi:10.1145/3290605.3300340}}.

\bibitem{polygone}
{Polygone Art}.
\newblock URL: \url{https://polygone.art/}.

\bibitem{rack2023extensible}
Christian Rack, Konstantin Kobs, Tamara Fernando, Andreas Hotho, and Marc~Erich Latoschik.
\newblock Extensible motion-based identification of {XR} users using non-specific motion data, 2023.
\newblock \href {https://arxiv.org/abs/2302.07517} {\path{arXiv:2302.07517}}.

\bibitem{beatleader}
Viktor Radulov.
\newblock {BeatLeader}.
\newblock URL: \url{https://www.beatleader.xyz/}.

\bibitem{blpolicy}
Viktor Radulov.
\newblock {BeatLeader} {Privacy} {Policy}.
\newblock URL: \url{https://www.beatleader.xyz/privacy}.

\bibitem{viewer}
Viktor Radulov, Kevin Ngo, Diego~F. Goberna, Buck Bukaty, James Kerrane, and Jack Baron.
\newblock Beat {Saber} web replays.
\newblock URL: \url{https://github.com/BeatLeader/BeatSaber-Web-Replays/graphs/contributors}.

\bibitem{cyberramen}
Dziugas Ramonas.
\newblock {CyberRamen}.
\newblock URL: \url{https://www.beatleader.xyz/u/165749}.

\bibitem{10.1145/2580723.2580730}
Franziska Roesner, Tadayoshi Kohno, and David Molnar.
\newblock Security and privacy for augmented reality systems.
\newblock {\em Commun. ACM}, 57(4):88–96, apr 2014.
\newblock \href {https://doi.org/10.1145/2580723.2580730} {\path{doi:10.1145/2580723.2580730}}.

\bibitem{schell_comparison_2022}
Christian Schell, Andreas Hotho, and Marc~Erich Latoschik.
\newblock Comparison of {Data} {Encodings} and {Machine} {Learning} {Architectures} for {User} {Identification} on {Arbitrary} {Motion} {Sequences}.
\newblock In {\em 2022 {IEEE} {International} {Conference} on {Artificial} {Intelligence} and {Virtual} {Reality} ({AIVR})}, pages 11--19, December 2022.
\newblock ISSN: 2771-7453.
\newblock \href {https://doi.org/10.1109/AIVR56993.2022.00010} {\path{doi:10.1109/AIVR56993.2022.00010}}.

\bibitem{scoresaber}
{ScoreSaber}.
\newblock URL: \url{https://www.scoresaber.com/}.

\bibitem{9833742}
Sophie Stephenson, Bijeeta Pal, Stephen Fan, Earlence Fernandes, Yuhang Zhao, and Rahul Chatterjee.
\newblock Sok: Authentication in augmented and virtual reality.
\newblock In {\em 2022 IEEE Symposium on Security and Privacy (SP)}, pages 267--284, 2022.
\newblock \href {https://doi.org/10.1109/SP46214.2022.9833742} {\path{doi:10.1109/SP46214.2022.9833742}}.

\bibitem{10027854}
Pier~Paolo Tricomi, Federica Nenna, Luca Pajola, Mauro Conti, and Luciano Gamberini.
\newblock You can’t hide behind your headset: User profiling in augmented and virtual reality.
\newblock {\em IEEE Access}, 11:9859--9875, 2023.
\newblock \href {https://doi.org/10.1109/ACCESS.2023.3240071} {\path{doi:10.1109/ACCESS.2023.3240071}}.

\bibitem{unity}
Unity {Real}-{Time} {Development} {Platform}: {3D}, {2D}, {VR}, and {AR} {Engine}.
\newblock URL: \url{https://unity.com}.

\bibitem{nist}
C.~L. Wilson.
\newblock Biometric accuracy standards, 2003.
\newblock URL: \url{https://csrc.nist.gov/CSRC/media/Events/ISPAB-MARCH-2003-MEETING/documents/March2003-Biometric-Accuracy-Standards.pdf}.

\bibitem{wobbeking_beat_2022}
Jan Wöbbeking.
\newblock Beat {Saber} generated more revenue in 2021 than the next five biggest apps combined, August 2022.
\newblock URL: \url{https://mixed-news.com/en/beat-saber-generated-more-revenue-in-2021-than-the-next-five-biggest-apps-combined/}.

\end{thebibliography}

\normalsize
\appendix
\section{Benchmarking}
\label{app:benchmarking}

For all experiments described in this paper, we used a desktop computer running Windows 10 v22H2 with 128 GB of 2133 MHz DDR4 RAM, an AMD Ryzen 9 5950X CPU (16 cores, 3.40 GHz), and an NVIDIA GeForce RTX 3090 GPU (10496 CUDA cores, 24 GB VRAM). The time required to run the experiments in each section was as follows:

\subsubsection*{Motivation (\S\ref{sec:motivation})}
\begin{itemize}[leftmargin=*]
    \itemsep 0em
    \item Preprocessing the BOXRR-23 dataset to sample and normalize 500 replays each for 500 users took \textbf{37h 14m}. 
    \item Training and testing the LSTM funnel models took \textbf{3h 50m}.
    \item Featurization for the Miller et al. \cite{miller_personal_2020} model took \textbf{2h 54m}, and training and testing the model took \textbf{4m 8s}.
    \item Featurization for the Nair et al. \cite{291259} model took \textbf{16h 32m}, and training and testing the model took \textbf{13m 47s}.
\end{itemize}

\subsubsection*{Method (\S\ref{sec:method})}
\begin{itemize}[leftmargin=*]
    \itemsep 0em
    \item Preprocessing the BOXRR-23 dataset to sample the action similarity features took \textbf{55h 40m}. Training and testing the action similarity model took \textbf{3h 26m}.
    \item Preprocessing the BOXRR-23 dataset to sample the user similarity features took \textbf{57h 12m}. Training and testing the user similarity model took \textbf{1h 18m}.
    \item Training and testing the anonymizer model took \textbf{3h 15m}.
    \item Training and testing the normalizer model took \textbf{56m 51s}.
\end{itemize}

\subsubsection*{Anonymity (\S\ref{sec:anonymity})}
\begin{itemize}[leftmargin=*]
    \itemsep 0em
    \item Preprocessing the BOXRR-23 dataset to sample and normalize 20 replays each for 1000 users took \textbf{3h 24m}.
    \item Featurization for the Miller et al. \cite{miller_personal_2020} model took \textbf{40m 51s}, and training and testing the model took \textbf{5m 9s}.
    \item Featurization for the Nair et al. \cite{291259} model took \textbf{3h 42m}, and training and testing the model took \textbf{1h 12m}.
    \item Training and testing the LSTM funnel model took \textbf{8m 52s}.
\end{itemize}

\subsubsection*{Interactivity (\S\ref{sec:interactivity})}
\begin{itemize}[leftmargin=*]
    \itemsep 0em
    \item Preprocessing the BOXRR-23 dataset to sample and anonymize 1,000 replays took \textbf{37m 10s}.
    \item Using SimSaber to simulate the 1,000 replays before and after anonymization took \textbf{4m 32s}.
\end{itemize}

\noindent Overall, the total compute time required was about \textbf{192h 52m}.

\eject
\clearpage
\onecolumn

\begin{figure}[ht]
    \section{Full Architecture}
    \label{app:architecture}
    \centering
    \includegraphics[width=\linewidth]{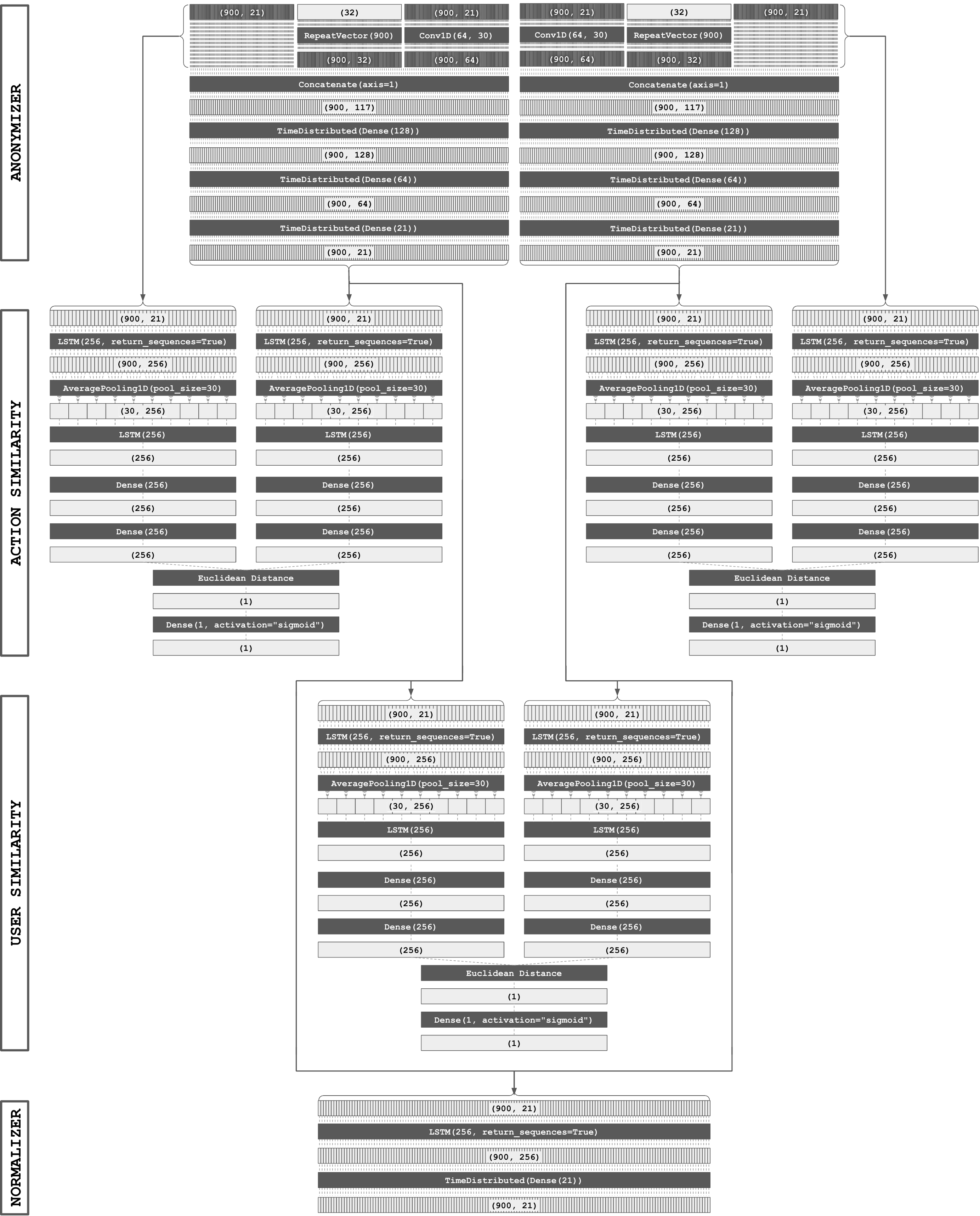}
\end{figure}

\end{document}